\documentclass[11pt]{article}
\usepackage{cite}
\usepackage{mdframed}
\usepackage{epsfig,subfigure}
\usepackage{amssymb}
\usepackage[fleqn]{amsmath}
\usepackage{color}
\usepackage{arydshln}
\def\QED{\mbox{\rule[0pt]{1.5ex}{1.5ex}}}
\def\proof{\noindent\hspace{2em}{\it Proof: }}

\usepackage{graphicx}
\usepackage{color}
\usepackage[dvipsnames]{xcolor}

\definecolor{armygreen}{rgb}{0.29, 0.33, 0.13}

\usepackage{amssymb}
\usepackage{amsmath,amsfonts,amssymb}
\usepackage{verbatim}
\usepackage{stfloats}
\usepackage[bookmarks=false]{}
\usepackage{algorithm}
\usepackage{algcompatible}
\usepackage{tikz,pgfplots,tikzpeople}

\newtheorem{theorem}{Theorem}
\newtheorem{corollary}{Corollary}
\newtheorem{definition}{Definition}
\newtheorem{lemma}{Lemma}

\newtheorem{remark}{Remark}

\newtheorem{example}{Example}

\newcommand\blfootnote[1]{%
  \begingroup
  \renewcommand\thefootnote{}\footnote{#1}%
  \addtocounter{footnote}{-1}%
  \endgroup
}

\begin{document}
\date{}

\title{
Weakly Secure Summation with Colluding Users
}
\author{\normalsize Zhou Li, Yizhou Zhao, Hua Sun \\
}

\maketitle

\blfootnote{
Zhou Li (email: zhouli@my.unt.edu), Yizhou Zhao (email: yizhouzhao@my.unt.edu), and Hua Sun (email: hua.sun@unt.edu) are with the Department of Electrical Engineering at the University of North Texas. }

\maketitle

\begin{abstract}
In secure summation, $K$ users, each holds an input, wish to compute the sum of the inputs at a server without revealing any information about {\em all the inputs} even if the server may collude with {\em an arbitrary subset of users}. In this work, we relax the security and colluding constraints, where the set of inputs whose information is prohibited from leakage is from a predetermined collection of sets (e.g., any set of up to $S$ inputs) and the set of colluding users is from another predetermined collection of sets (e.g., any set of up to $T$ users). For arbitrary collection of security input sets and colluding user sets, we characterize the optimal randomness assumption, i.e., the minimum number of key bits that need to be held by the users, per input bit, for weakly secure summation to be feasible, which generally involves solving a linear program.
\end{abstract}

\newpage

\allowdisplaybreaks
\section{Introduction}
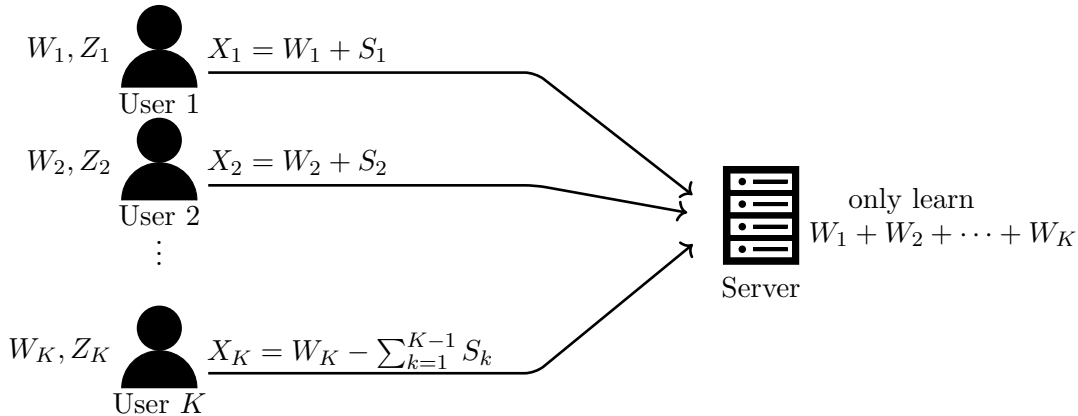
\begin{figure}[h]
\centering
\begin{tikzpicture}
    \node (u1) at (0,4) {};
    \node (u2) at (0,2.5) {};
    \node at (0.5,1.9) {$\vdots$};
    \node (uK) at (0,0) {};
    \node (server) at (8,2.25) {};
    \node (server1) at ($(server)+(0.1,-0.5)$) {};
    \node (server2) at ($(server)+(0.1,-0.2)$) {};
    \node (server3) at ($(server)+(0.1,0.1)$) {};
    \node (server4) at ($(server)+(0.1,0.4)$) {};
    \filldraw ($(u1)$)
    to[out=90,in=180] ($(u1) + (0.5,0.5)$)
    to[out=0,in=90] ($(u1) + (1,0)$);
    \fill ($(u1) + (0.5,0.8)$) circle(0.3);
    \filldraw ($(u2)$)
    to[out=90,in=180] ($(u2) + (0.5,0.5)$)
    to[out=0,in=90] ($(u2) + (1,0)$);
    \fill ($(u2) + (0.5,0.8)$) circle(0.3);
    \filldraw ($(uK)$)
    to[out=90,in=180] ($(uK) + (0.5,0.5)$)
    to[out=0,in=90] ($(uK) + (1,0)$);
    \fill ($(uK) + (0.5,0.8)$) circle(0.3);
    \filldraw ($(server)+(0,-0.6)$) rectangle ($(server)+(1,0.7)$);
    \foreach \v in {1,2,...,4} {
        \filldraw [white] (server\v) rectangle ($(server\v)+(0.8,0.2)$);
        \filldraw ($(server\v)+(0.3,0.08)$) rectangle ($(server\v)+(0.75,0.12)$);
        \fill ($(server\v)+(0.15,0.1)$) circle (0.05);
    }
    \draw [rounded corners,->, line width=1pt,shorten >=10pt]($(u1) + (1.15,0.2)$) -- ($(u1) + (5.5,0.2)$)
    -- (server);
    \draw [rounded corners,->, line width=1pt,shorten >=10pt]($(u2) + (1.15,0.2)$) -- ($(u2) + (5.5,0.2)$)
    -- (server);
    \draw [rounded corners,->, line width=1pt,shorten >=10pt]($(uK) + (1.15,0.2)$) -- ($(uK) + (5.5,0.2)$)
    -- (server);
    \node at ($(u1) + (1,0.5)$) [right]{$X_1=W_1+S_1$};
    \node at ($(u2) + (1,0.5)$) [right]{$X_2=W_2+S_2$};
    \node at ($(uK) + (1,0.5)$) [right]{$X_K=W_K-\sum_{k=1}^{K-1} S_k$};
    \node at ($(u1) + (0.5,-0.2)$) {User $1$};
    \node at ($(u2) + (0.5,-0.2)$) {User $2$};
    \node at ($(uK) + (0.5,-0.2)$) {User $K$};
    \node at ($(u1) + (0,0.5)$) [left]{$W_1,Z_1$};
    \node at ($(u2) + (0,0.5)$) [left]{$W_2,Z_2$};
    \node at ($(uK) + (0,0.5)$) [left]{$W_K,Z_K$};
    \node at ($(server)+(0.5,-0.9)$) []{Server};
    \node at ($(server)+(2.5,0.25)$) []{only learn};
    \node at ($(server)+(1,-0.2)$) [right]{$W_1+W_2+\cdots+W_K$};
\end{tikzpicture}
\caption{The secure summation problem and an optimal protocol where $S_1, \cdots, S_{K-1}$ are uniform and independent (and determine $Z_k$ as $Z_k = S_k$, where $k \in \{ 1,\cdots, K-1 \}$, and $Z_K = -\sum_{k=1}^{K-1} S_k$).}
\label{fig:model}
\end{figure}

The focus of this work is on the information theoretic secure summation problem \cite{Zhao_Sun_Summation} (see Fig.~\ref{fig:model}), where User $k \in \{1,2,\cdots,K\}$ holds an input variable $W_k$ and an independent key variable $Z_k$ from a finite field, and is connected to a server through a noiseless orthogonal link. From one message $X_k$ from each user, the server shall be able to decode the sum of the inputs $W_1+\cdots+W_K$ while obtaining no additional information about all the inputs $W_1, \cdots, W_K$.

An optimal secure summation protocol is plotted in Fig.~\ref{fig:model}, where the key variables are $(K-1)$-MDS and zero-sum, i.e., any $K-1$ variables from $Z_1, \cdots, Z_K$ are independent and uniform, and $Z_1 + \cdots + Z_K$ is $0$. The optimality of the protocol is regarding both the communication cost and the randomness cost, i.e., in order to compute $1$ bit of the summation securely, each user must send a message $X_k$ of at least $1$ bit to the server and the $K$ users need to hold key variables of at least $K-1$ bits (the joint entropy of $Z_1,\cdots, Z_K$). Note that the randomness cost scales linearly with the total number of users $K$, which could be huge in practice. This is mainly due to the stringent security constraint, i.e., we wish to protect all $K$ inputs. One main motivation of this work is to relax the security constraint to a weaker one (i.e., the set of inputs that need to be kept secure are some subsets of all inputs) and understand its impact on the randomness cost. Moreover, the minimum communication cost and randomness cost for secure summation remain unchanged even if user-server collusion is included \cite{Zhao_Sun_Summation}. In particular, no matter which set of users (big or small) may collude with the server so that the server might get some advantage in inferring information about the remaining users, the optimal protocol remains the same. The other main motivation of this work is to see if the dependence on the colluding pattern will be more explicit in the weakly secure summation problem, i.e., we wish to study the joint effect of arbitrary security and colluding patterns on the randomness consumption.

The main result of this work is a complete characterization of the minimum key size for weakly secure summation with arbitrary security and colluding patterns, i.e., arbitrary security input sets and colluding user sets. The ultimate answer generally involves two parts - one integral part that corresponds to the number of users that need to be protected under a pair of security input set and colluding user set and one possibly fractional part that corresponds to the amount of key required for remaining users whose value is determined by a linear program.

\section{Problem Statement and Definitions}\label{sec:model}
Consider one server and $K \geq 2$ users, where 
User $k \in \{1,2,\cdots,K\} \triangleq [K]$ holds an input vector $W_k$ and a key variable $Z_k$. The inputs $\left(W_k\right)_{k\in[K]}$ are independent. Each $W_k$ is an $L \times 1$ column vector and the $L$ elements are i.i.d. uniform symbols from the finite field $\mathbb{F}_q$. $\left(W_k\right)_{k\in[K]}$ is independent of $\left(Z_k\right)_{k\in[K]}$. 
\begin{eqnarray}
   && H\left(\left(W_k\right)_{k\in[K]},
    \left(Z_k\right)_{k\in[K]}\right)=
    \sum_{k\in[K]} H\left(W_k \right) +
    H\left(\left(Z_k\right)_{k\in[K]} \right), \label{ind} \\
   && H(W_k) = L ~(\mbox{in $q$-ary units}), ~\forall k \in [K]. \label{h2}
\end{eqnarray}
The key variables can be arbitrarily correlated and are a function of a source key variable $Z_\Sigma$, which is comprised of $L_{Z_{\Sigma}}$ symbols from $\mathbb{F}_q$.
\begin{eqnarray}
	H\left(\left(Z_k\right)_{k\in[K]} \Big| Z_\Sigma\right) = 0.\label{total rand}
\end{eqnarray}
User $k$ sends to the server a message $X_k$, which is a function of $W_k, Z_k$ and consists of $L_X$ symbols from $\mathbb{F}_q$.
\begin{eqnarray}
    H\left(X_k | W_k, Z_k\right) = 0, \forall k \in [K].\label{message}
\end{eqnarray}
From all messages, the server must be able to recover the desired sum $\sum_{k \in [K]} W_k$ with no error.
\begin{eqnarray}
    \mbox{[Correctness]}~~~H\left(\sum_{k\in[K]} W_k \Bigg| \left(X_k\right)_{k\in[K]} \right) = 0.\label{corr}
\end{eqnarray}
The security input sets are described by a monotone\footnote{A set system is called monotone if a set belongs to the system, then its subset also belongs to the system.} set system $\{\mathcal{S}_1, \cdots, \mathcal{S}_M\}$ and the colluding user sets are described by another monotone set\footnote{Without loss of generality, assume $\cup_m \mathcal{S}_m \neq \emptyset$ (the security constraints are not empty) and $|\mathcal{T}_n|\leq K-2$ as otherwise there is nothing to hide.} system $\{\mathcal{T}_1,\cdots, \mathcal{T}_N\}$. The security constraint states that if the server colludes with users from any $\mathcal{T}_n$ set, nothing is revealed about the inputs from any $\mathcal{S}_m$ set (excluding what is possibly already known to users from $\mathcal{T}_n$ when $\mathcal{S}_m \cap \mathcal{T}_n \neq \emptyset$),
\begin{eqnarray}
\mbox{[Security]}~~~I\left(\left(W_k\right)_{k\in\mathcal{S}_m}; \left(X_k\right)_{k\in[K]} \Bigg| \sum_{k\in [K]} W_k, \left( W_k, Z_k \right)_{k\in\mathcal{T}_n} \right) = 0, ~\forall m \in [M], n \in [N].\label{security}
\end{eqnarray}
The 
key {\em rate} $R_{Z_{\Sigma}}$, characterizes how many symbols the source key variable contains per input symbol, and is defined as follows.
\begin{eqnarray}
    R_{Z_{\Sigma}} \triangleq \frac {L_{Z_{\Sigma}}}{L}. \label{rate:R_Z_sum}
\end{eqnarray}
The rate $R_{Z_{\Sigma}}$ is said to be achievable if there exists a secure summation scheme, for which the correctness constraint (\ref{corr}) and the security constraint (\ref{security}) are satisfied, and the key rate is no greater than $R_{Z_{\Sigma}}$. The infimum of the achievable rates $R_{Z_{\Sigma}}$ is called the optimal key rate, denoted as $R_{Z_{\Sigma}}^*$.

\subsection{Auxiliary Definitions}
To facilitate the presentation of our results, we introduce some auxiliary definitions in this section. 

Some users are imposed to be protected by some key variable even if they do not explicitly belong to any $\mathcal{S}_m$ set. Such implicit security sets are specified below. For two sets $\mathcal{A}, \mathcal{B}$, the set difference $\mathcal{A} \setminus \mathcal{B}$ is defined as the set of elements that belong to $\mathcal{A}$ but not to $\mathcal{B}$.

\begin{definition}[Implicit Security Input Set $\mathcal{S}_I$] \label{def:imp} The implicit security input set is defined as 
\begin{eqnarray}
\mathcal{S}_I\triangleq \Big\{ [K]\setminus\{\mathcal{S}_m\cup\mathcal{T}_n\} : |\mathcal{S}_m\cup\mathcal{T}_n|=K-1, \forall m \in [M], \forall n \in [N] \Big\} \setminus \{\cup_{i\in [M]}\mathcal{S}_i \}.
\end{eqnarray}
\end{definition}

We will use the following example to explain the definitions.
\begin{example}\label{ex1}
Consider $K=5$, the security input sets are $(\mathcal{S}_1,\cdots,\mathcal{S}_4)=(\emptyset, \{1\}, \{2\}, \{3\})$, and the colluding user sets are $(\mathcal{T}_1,\cdots,\mathcal{T}_{14})=(\emptyset, \{1\}, \{2\}, \{3\}, \{4\},$ $\{5\}, \{1,3\}, \{1,4\}, \{2,3\}, \{2,5\}, \{3,4\}$, $\{3,5\}$, $\{1,3,4\}, \{2,3,5\})$.
\end{example}

Searching for all security input set $\mathcal{S}_m$ and colluding user set $\mathcal{T}_n$ whose union has cardinality $K-1 = 4$, we have $|\mathcal{S}_2\cup\mathcal{T}_{14}|=|\{1\}\cup\{2,3,5\}|=4$ and $|\mathcal{S}_3\cup\mathcal{T}_{13}|=|\{2\}\cup\{1,3,4\}|=4$, so $\mathcal{S}_I=\{4,5\}$ for Example \ref{ex1}.

\begin{definition}[Total Security Input Set $\overline{\mathcal{S}}$] \label{def:tot} 
The union of explicit and implicit security input sets is defined as the total security input set, 
\begin{eqnarray}
\overline{\mathcal{S}}\triangleq \cup_{m\in[M]} \mathcal{S}_m 
\cup\mathcal{S}_I.
\end{eqnarray}
\end{definition}

For Example \ref{ex1}, we have $\overline{\mathcal{S}}=\{1\} \cup \{2\} \cup \{3\}\cup\{4,5\}=\{1,2,3,4,5\}$.

\begin{definition}[Intersection of $\mathcal{S}_m \cup \mathcal{T}_n$ and $\overline{\mathcal{S}}$, $\mathcal{A}_{m,n}$] \label{def:totset}
For each pair of security input set $\mathcal{S}_m$ and colluding user set $\mathcal{T}_n$,  
its overlap with the total security input set $\overline{\mathcal{S}}$ is denoted as
\begin{eqnarray}
\mathcal{A}_{m,n}\triangleq (\mathcal{S}_m\cup\mathcal{T}_n)\cap\overline{\mathcal{S}} 
\end{eqnarray}
and its maximum cardinality is denoted as 
\begin{eqnarray}
a^*\triangleq \max_{m\in[M], n \in [N]} |\mathcal{A}_{m,n}|. 
\end{eqnarray}
\end{definition}
For Example \ref{ex1}, $\mathcal{A}_{2,3}= (\{1\}\cup\{2\})\cap\{1,2,3,4,5\} = \{1,2\}$, 
$\mathcal{A}_{2,14}= (\{1\}\cup\{2,3,5\})\cap\{1,2,3,4,5\} = \{1,2,3,5\}$, and $\mathcal{A}_{3,13}= (\{2\}\cup\{1,3,4\})\cap\{1,2,3,4,5\} = \{1,2,3,4\}$. Further, $a^*=4$.

\begin{definition}[Union of Maximum $\mathcal{A}_{m,n}$] \label{def:uni}
Find all $\mathcal{A}_{m,n}$ sets with maximum cardinality
and denote the union of the corresponding $\mathcal{S}_m, \mathcal{T}_n$ sets as $\mathcal{Q}$.
\begin{eqnarray}
\mathcal{Q} \triangleq \cup_{m,n: |\mathcal{A}_{m,n}| = a^*} \mathcal{S}_m \cup \mathcal{T}_n. 
\end{eqnarray}
\end{definition}
For Example \ref{ex1}, $\mathcal{A}_{2,14}, \mathcal{A}_{3,13}$ are all $\mathcal{A}_{m,n}$ sets with the maximum cardinality, so 
$\mathcal{Q} = \mathcal{S}_2 \cup \mathcal{T}_{14} \cup \mathcal{S}_{3} \cup \mathcal{T}_{13} 
= \{1,2,3,4,5\}$.

\section{Result}
\begin{theorem}\label{thm:d1}
For secure summation with $K \geq 2$ users, security input sets $(\mathcal{S}_m)_{m\in[M]}$, and colluding user sets $(\mathcal{T}_n)_{n\in[N]}$, 
the optimal key rate $R_{Z_{\Sigma}}^*$ is 
\begin{eqnarray}
    R_{Z_{\Sigma}}^* = \left\{
    \begin{array}{cl}
        a^* +b^*& ~\mbox{if}~a^*\leq K-1, a^*=\big|\overline{\mathcal{S}}\big|, ~\mbox{and}~|\mathcal{Q}|=K\\
         \min(a^*,K-1)& ~\mbox{otherwise}
    \end{array}
    \right.
\end{eqnarray}
where $b^*$ is the optimal value of the following linear program\footnote{Note that for the `if' case, $|\mathcal{S}_m \cup \mathcal{T}_n| \leq K-1$ because otherwise $a^* = K$; for all $m,n$ such that $|\mathcal{A}_{m,n}| = a^*$, $a^* = |\overline{\mathcal{S}}|$ so $\overline{\mathcal{S}} \subset (\mathcal{S}_m \cup \mathcal{T}_n)$ and $b_k, k \in [K] \setminus \overline{\mathcal{S}}$ are all the variables.}, 
 \begin{eqnarray}
&& \min \max_{m,n: |\mathcal{A}_{m,n}| = a^* }  \sum_{k\in \mathcal{T}_{n}\setminus\overline{\mathcal{S}}} b_{k} \notag
\\
 \text{subject to} 
    && \sum_{k\in [K]\setminus (\mathcal{S}_m \cup \mathcal{T}_n)} b_{k} 
    \geq 1,\forall m,n ~\mbox{such that}~ |\mathcal{A}_{m,n}| = a^*, 
    \\
    && b_{k}\geq 0, \forall k\in [K]\setminus\overline{\mathcal{S}}.
    \label{lp}
 \end{eqnarray}
\end{theorem}

Theorem \ref{thm:d1} applies to arbitrary security input sets $\mathcal{S}_m$ and arbitrary colluding user sets $\mathcal{T}_n$, and for some cases the optimal rate value may not be immediately seen. For example, consider the symmetric case, where $\mathcal{S}_m$ contains all subsets of $[K]$ with cardinality at most $S$ and $\mathcal{T}_n$ contains all subsets of $[K]$ with cardinality at most $T$, then $R_{Z_{\Sigma}}^* = \min(S+T, K-1)$. Note that for the symmetric case, $\overline{\mathcal{S}} = [K]$ and we never fall into the `if' case in Theorem \ref{thm:d1}. The true power of Theorem \ref{thm:d1} lies in the more heterogeneous case where `if' part comes into play.

\section{Converse Proof of Theorem \ref{thm:d1}} \label{sec:converse}
We start from the `otherwise' case and show that $R_{Z_\Sigma} \geq \min(a^*, K-1)$. Let's use Example \ref{ex1} to illustrate the idea.
\subsection{Proof of Example \ref{ex1}} \label{sec:ex1}

The converse proof is based on showing that for any $\mathcal{A}_{m,n}$, we have $H\left( \left(Z_k\right)_{k \in \mathcal{A}_{m,n}} \right) \geq |\mathcal{A}_{m,n}| L$.

For Example \ref{ex1}, let's take $\mathcal{A}_{3,13} = \{1,2,3,4\}$ as an example, where $\{4\}$ comes from the implicit security input set $\mathcal{S}_I$ and $\{1,2,3\}$ comes from the explicit security input set $\cup_m \mathcal{S}_m$. When expanding $H\left( \left(Z_k\right)_{k \in \mathcal{A}_{m,n}} \right)$, we first consider the term from the implicit set and then consider the term from the explicit set (conditioned on the implicit set).
\begin{eqnarray}
H(Z_1,Z_2,Z_3,Z_4)
&=&H(Z_4)+H(Z_1,Z_2,Z_3|Z_4)
\end{eqnarray}
and next, we proceed to show that $H(Z_4) \geq L$ and $H(Z_1,Z_2,Z_3|Z_4) \geq 3L$.

First, consider $H(Z_4) \geq L$. The intuition is that $\{4\} = [K] \setminus (\mathcal{S}_2\cup\mathcal{T}_{14}) = [5]\setminus(\{1\}\cup\{2,3,5\})$ belongs to the implicit security input set, i.e., when the server colludes with users in $\mathcal{T}_{14}$, the sum $\sum_{k\in[K]} W_k$ can be decoded and nothing is revealed about $\left(W_k\right)_{k\in\mathcal{S}_2} $. Complementarily, nothing shall be revealed about $\left(W_k\right)_{k\in [K] \setminus (\mathcal{S}_2\cup\mathcal{T}_{14})} = W_4$. 
Expressing this idea in entropy terms, we have
\begin{eqnarray}
&&H(Z_4) \geq H\left(Z_{4}\big|(Z_k)_{k\in \mathcal{T}_{14}}\right) = H(Z_4|Z_2,Z_3,Z_5)\\
&\geq&I(Z_4;Z_1|Z_2,Z_3,Z_5) \overset{(\ref{ind})}{=} I(Z_4,W_4;Z_1,W_1|Z_2,Z_3,Z_5,W_2,W_3,W_5) \label{eq:t1}\\
&\overset{(\ref{message})}{\geq}&I(X_4,W_4;X_1,W_1|Z_2,Z_3,Z_5,W_2,W_3,W_5) \geq I(X_4,W_4;W_1|Z_2,Z_3,Z_5,W_2,W_3,W_5,X_1) \notag\\
&&\label{eq:t2}\\
&=&H(W_1|Z_2,Z_3,Z_5,W_2,W_3,W_5,X_1)-H(W_1|Z_2,Z_3,Z_5,W_2,W_3,W_5,X_1,X_4,W_4)\\
&\overset{(\ref{message})(\ref{corr})}{\geq}&H\left(W_1\Bigg|Z_2,Z_3,Z_5,W_2,W_3,W_5, (X_k)_{k\in[5]}, \sum_{k\in[5]} W_k\right) \notag\\
&&-~H\left(W_1\Bigg|Z_2,Z_3,Z_5,W_2,W_3,W_5,X_2, X_3, X_5, X_1,X_4,W_4, \sum_{k\in[5]} W_k\right) \label{eq:t3}\\
&\overset{(\ref{security})}{\geq}&H\left(W_1\Bigg|Z_2,Z_3,Z_5,W_2,W_3,W_5,\sum_{k\in[5]} W_k\right)-H(W_1|W_1) \label{eq:t4}\\
&\overset{(\ref{ind})(\ref{h2})}{=}&L - 0 = L \label{impsec1}
\end{eqnarray}
where (\ref{eq:t1}) follows from the independence of $(W_k)_{k\in[K]}$ and $(Z_k)_{k\in[K]}$ (refer to (\ref{ind})) and (\ref{eq:t2}) follows from the fact that $X_k$ is a function of $W_k, Z_k$ (refer to (\ref{message})). In (\ref{eq:t3}), to obtain the first term, we have added conditioning on all $X_k$ and the desired sum $\sum_k W_k$; to obtain the second term, we use the correctness constraint (\ref{corr}) that from all $X_k$, we can decode $\sum_k W_k$. In (\ref{eq:t4}), the first term follows from the security constraint (\ref{security}) where the security input set is $\mathcal{S}_2 = \{1\}$ and the colluding user set is $\mathcal{T}_{14} = \{2,3,5\}$; the second term follows from dropping all other terms except $W_1$ (obtained from $\sum_k W_k$ and $W_2, W_3, W_5, W_4$). The last step follows from the independence of $(W_k)_{k\in[K]}$ and $(Z_k)_{k\in[K]}$ (refer to (\ref{ind})) and the uniformity of $(W_k)_{k\in[K]}$ (refer to (\ref{h2})).

Second, consider $H(Z_1,Z_2,Z_3|Z_4) \geq 3L$. Note that $\mathcal{S}_3\cup\mathcal{T}_{13}=\{2\}\cup\{1,3,4\}=\{1,2,3,4\}$ so that $Z_4$ may appear in the conditioning term and $(\mathcal{S}_3\cup\mathcal{T}_{13})\cap\cup_{m\in [4]}\mathcal{S}_m=\{1,2,3,4\}\cap\{1,2,3\}=\{1,2,3\}$ so that we want to show that $Z_1,Z_2,Z_3$ must each contribute $L$ independent amount of information.
\begin{eqnarray}
&& H(Z_1,Z_2,Z_3|Z_4)\notag\\
&\geq&H(Z_1,Z_2,Z_3|Z_4,W_1,W_2,W_3,W_4)\\
&\geq&I(Z_1,Z_2,Z_3;X_1,X_2,X_3|Z_4,W_1,W_2,W_3,W_4)\\
&\overset{(\ref{message})}{=}&H(X_1,X_2,X_3|Z_4,W_1,W_2,W_3,W_4)\\
&=&H(X_1,X_2,X_3|W_4,Z_4)-I(X_1,X_2,X_3;W_1,W_2,W_3|W_4,Z_4)\\
&\overset{}{\geq}&3L- I(X_1,X_2,X_3;W_1|W_4,Z_4) - I(X_1,X_2,X_3;W_3|W_4,Z_4,W_1) \notag\\
&&-~I(X_1,X_2,X_3;W_2|W_4,Z_4,W_1,W_3) \label{sec_indi}\\
&\overset{}{\geq}& 3L- I\left(X_1,X_2,X_3, \sum_{k\in[5]} W_k ;W_1\Bigg|W_4, Z_4\right) - I\left(X_1,X_2,X_3,\sum_{k\in[5]} W_k, Z_1;W_3\Bigg|W_4,Z_4,W_1\right) \notag\\
&&-~I\left(X_1,X_2,X_3,\sum_{k\in[5]} W_k, Z_1, Z_3;W_2\Bigg|W_4,Z_4,W_1,W_3\right)\\
&\overset{(\ref{ind})}{=}& 3L- I\left(X_1,X_2,X_3 ;W_1\Bigg|\sum_{k\in[5]} W_k, W_4, Z_4\right) - I\left(X_1,X_2,X_3;W_3\Bigg|\sum_{k\in[5]} W_k, W_4,Z_4,W_1, Z_1\right) \notag\\
&&-~I\left(X_1,X_2,X_3;W_2\Bigg|\sum_{k\in[5]} W_k, W_4,Z_4,W_1,Z_1, W_3, Z_3\right) \label{totalsec1} \\
&\overset{(\ref{security})}{=}&3L 
\end{eqnarray}
where in (\ref{sec_indi}), $H(X_1,X_2,X_3|W_4,Z_4) \geq 3L$ follows from the transmission size constraint, which will be proved in Lemma \ref{lemma1} and the remaining terms follow from applying security constraints for various security input set and colluding user set. Specifically, in (\ref{totalsec1}), the second term is  zero due to the security constraint (\ref{security}) with $\mathcal{S}_2=\{1\}$ and $\mathcal{T}_5=\{4\}$, the third term is zero due to the security constraint (\ref{security}) with $\mathcal{S}_4=\{3\}$ and $\mathcal{T}_8=\{1,4\}$ and the fourth term is zero due to the security constraint (\ref{security}) with $\mathcal{S}_3=\{2\}$ and $\mathcal{T}_{13}=\{1,3,4\}$. Note that the order of chain-rule expansion is carefully chosen, where $W_2$ is considered last as it belongs to $\mathcal{S}_3$ while the other terms $W_1, W_3$ are considered first as they belong to $\mathcal{T}_{13}$ and the set systems are monotone (remember that the term of consideration $H(Z_1, Z_2, Z_3, Z_4)$ comes from $\mathcal{A}_{3,13}$).

\subsection{Proof of $R_{Z_\Sigma} \geq \min(a^*, K-1)$}
We are now ready to generalize the above proof to all parameter settings. Let's start with a few useful lemmas. 
First as a preparation, we show that each $X_k$ must contain at least $L$ symbols (the size of the input) even if all other inputs are known.
This result has appeared as Lemma 1 in \cite{Zhao_Sun_Summation} and a proof is presented here for completeness.
\begin{lemma}\label{lemma1}
For any $u \in [K]$, we have 
\begin{eqnarray}
    H\left(X_u\big|(W_k,Z_k)_{k\in[K]\backslash \{u\}}\right) \geq L. \label{lemma1_eq}
\end{eqnarray}
\end{lemma}

\proof
\begin{eqnarray}
    && H\left(X_u\big|(W_k,Z_k)_{k\in[K]\backslash \{u\}}\right)\notag\\
    &\geq& I\left(X_u; \sum_{k\in[K]} W_k \Bigg|(W_k,Z_k)_{k\in[K]\backslash \{u\}}\right)\\
    &=& H\left(\sum_{k\in[K]} W_k \Bigg| (W_k,Z_k)_{k\in[K]\backslash \{u\}}\right) - H\left(\sum_{k\in[K]} W_k \Bigg| X_u, (W_k,Z_k)_{k\in[K]\backslash \{u\}}\right)\\
    &\overset{(\ref{ind})(\ref{message})}{\geq}& H\left(W_u\right) - 
    H\left(\sum_{k\in[K]} W_k \Bigg| (X_k)_{k\in[K]}\right)
    \label{pf_lemma1_1}\\
    &\overset{(\ref{h2})(\ref{corr})}{=}& L \label{eq:e1} 
\end{eqnarray}
where the first term of (\ref{pf_lemma1_1}) follows from the fact that input $W_u$ is independent of other inputs and keys $(W_k,Z_k)_{k \in [K]\backslash\{u\}}$ (see (\ref{ind})) and the second term of (\ref{pf_lemma1_1}) follows from the fact that $(X_k)_{k\in[K]\backslash \{u\}}$ is determined by $(W_k,Z_k)_{k\in[K]\backslash \{u\}}$ (see (\ref{message})). In (\ref{eq:e1}), we use the property that $W_u$ has $L$ uniform symbols (see (\ref{h2})) and the desired sum $\sum_{k\in[K]} W_k$ can be decoded with no error from all messages $(X_k)_{k\in[K]}$ (see (\ref{corr})).

\hfill\QED

Next, we show that due to the security constraint, the keys used by users outside $\mathcal{S}_m\cup\mathcal{T}_n$
should not be less than $L$ symbols, conditioned on what is known by the colluding users.
\begin{lemma} \label{lemma2} 
For any $\mathcal{S}_m, \mathcal{T}_n, m\in [M],n\in [N]$ such that $\mathcal{S}_m \cap \mathcal{T}_n = \emptyset$ and $|\mathcal{S}_m\cup\mathcal{T}_n| \leq K-1$, we have 
\begin{eqnarray}
    H\left((Z_k)_{k \in [K] \setminus(\mathcal{S}_m\cup\mathcal{T}_n)}\big|(Z_k)_{k\in\mathcal{T}_n}\right) \geq L. \label{lemma2_eq}
\end{eqnarray}
\end{lemma}

\proof
\begin{eqnarray}
&&H\left((Z_k)_{k \in [K] \setminus(\mathcal{S}_m\cup\mathcal{T}_n)}\big|(Z_k)_{k\in\mathcal{T}_n}\right) \notag\\
&\geq&I\left((Z_k)_{k \in [K] \setminus(\mathcal{S}_m\cup\mathcal{T}_n)};(Z_k)_{k \in\mathcal{S}_m}\big|(Z_k)_{k\in\mathcal{T}_n}\right)\\
&\overset{(\ref{ind})}{=}&I\left((Z_k,W_k)_{k \in [K] \setminus(\mathcal{S}_m\cup\mathcal{T}_n)};(Z_k,W_k)_{k \in\mathcal{S}_m}\big|(Z_k,W_k)_{k\in\mathcal{T}_n}\right) \label{eq:tx1}\\
&\overset{(\ref{message})}{\geq}&I\left((X_k,W_k)_{k \in [K] \setminus(\mathcal{S}_m\cup\mathcal{T}_n)};(X_k,W_k)_{k \in\mathcal{S}_m}\big|(Z_k,W_k)_{k\in\mathcal{T}_n}\right)\\
&\geq&I\left((X_k,W_k)_{k \in [K] \setminus(\mathcal{S}_m\cup\mathcal{T}_n)};(W_k)_{k \in\mathcal{S}_m}|(Z_k,W_k)_{k\in\mathcal{T}_n},(X_k)_{k \in\mathcal{S}_m}\right)\\
&=&H\left((W_k)_{k \in\mathcal{S}_m}\big|(Z_k,W_k)_{k\in\mathcal{T}_n},(X_k)_{k \in\mathcal{S}_m}\right)\notag \\
&&-~H\left((W_k)_{k \in\mathcal{S}_m}\big|(Z_k,W_k)_{k\in\mathcal{T}_n},(X_k)_{k \in\mathcal{S}_m},(X_k,W_k)_{k \in [K] \setminus(\mathcal{S}_m\cup\mathcal{T}_n)}\right)\\
&\overset{(\ref{message})(\ref{corr})}{\geq}&H\left((W_k)_{k \in\mathcal{S}_m}\Bigg|\sum_{k\in [K]} W_k, \left( W_k, Z_k \right)_{k\in\mathcal{T}_n}, (X_k)_{k \in[K]} \right)-H\left((W_k)_{k \in\mathcal{S}_m}\Bigg|\sum_{k\in \mathcal{S}_m} W_k\right)\\
&\overset{(\ref{security})}{=}&H\left((W_k)_{k \in\mathcal{S}_m}\Bigg|\sum_{k\in [K]} W_k, \left( W_k, Z_k \right)_{k\in\mathcal{T}_n}\right)-H\left((W_k)_{k \in\mathcal{S}_m}\Bigg|\sum_{k\in \mathcal{S}_m} W_k\right)\\
&\overset{(\ref{ind})(\ref{h2})}{=}&|\mathcal{S}_m| L-(|\mathcal{S}_m|-1) L \label{eq:tt2}\\
&=&L
\end{eqnarray}
where a more detailed proof of (\ref{eq:tx1}) can be found in Lemma 5 of \cite{Zhao_Sun_MDS} and the first term of (\ref{eq:tt2}) uses the fact that $\mathcal{S}_m \cap \mathcal{T}_n = \emptyset$ and $|\mathcal{S}_m\cup\mathcal{T}_n| \leq K-1$.

\hfill\QED

Specializing Lemma \ref{lemma2} to the case where $|\mathcal{S}_m\cup\mathcal{T}_n| = K-1$, we have the following corollary for members from the implicit security input set $\mathcal{S}_{I}$ (refer to Definition \ref{def:imp}). 
\begin{corollary}
For any $\mathcal{S}_m, \mathcal{T}_n, m\in [M],n\in [N]$ such that $\mathcal{S}_m \cap \mathcal{T}_n = \emptyset$ and $|\mathcal{S}_m\cup\mathcal{T}_n| = K-1$, denote $u = [K]\setminus (\mathcal{S}_m\cup\mathcal{T}_n)$ and we have 
\begin{eqnarray}
    H\left(Z_u\big|(Z_k)_{k\in\mathcal{T}_n}\right) \geq L. \label{corollary1_eq}
\end{eqnarray}
\end{corollary}

We now proceed to consider any $\mathcal{S}_m, \mathcal{T}_n$ and show that the keys used by users in the explicit security input sets $(\mathcal{S}_m\cup\mathcal{T}_n)\cap(\cup_{i\in [M]} \mathcal{S}_i)$ must be at least its cardinality times $L$, conditioned on what is known by colluding users.
\begin{lemma} \label{lemma3} 
For any $\mathcal{S}_m, \mathcal{T}_n, m\in [M],n\in [N]$ such that $\mathcal{S}_m \cap \mathcal{T}_n = \emptyset$ and $|\mathcal{S}_m\cup\mathcal{T}_n| \leq K-1$, we have 
\begin{eqnarray}
    H\left((Z_k)_{k \in (\mathcal{S}_m\cup\mathcal{T}_n)\cap(\cup_{i\in [M]} \mathcal{S}_i)}\big|(Z_k)_{k\in\mathcal{T}_n\setminus (\cup_{i\in [M]} \mathcal{S}_i)}\right) \geq |(\mathcal{S}_m\cup\mathcal{T}_n)\cap(\cup_{i\in [M]}\mathcal{S}_i)| L. \label{lemma3_eq}
\end{eqnarray}
\end{lemma}

\proof
\begin{eqnarray}
&&H\left((Z_k)_{k \in (\mathcal{S}_m\cup\mathcal{T}_n)\cap(\cup_{i\in [M]} \mathcal{S}_i)}\big|(Z_k)_{k\in\mathcal{T}_n\setminus (\cup_{i\in [M]} \mathcal{S}_i)}\right) \notag\\
&\geq&H\left((Z_k)_{k \in (\mathcal{S}_m\cup\mathcal{T}_n)\cap(\cup_{i\in [M]} \mathcal{S}_i)}\big|(Z_k)_{k\in\mathcal{T}_n\setminus (\cup_{i\in [M]} \mathcal{S}_i)},(W_k)_{k \in (\mathcal{S}_m\cup\mathcal{T}_n)\cap(\cup_{i\in [M]} \mathcal{S}_i)}\right) \\
&\geq&I\left((Z_k)_{k \in (\mathcal{S}_m\cup\mathcal{T}_n)\cap(\cup_{i\in [M]} \mathcal{S}_i)};(X_k)_{k \in (\mathcal{S}_m\cup\mathcal{T}_n)\cap(\cup_{i\in [M]} \mathcal{S}_i)}\right.\big| \cdots \notag\\
&&\cdots \left.(Z_k)_{k\in\mathcal{T}_n\setminus (\cup_{i\in [M]} \mathcal{S}_i)},(W_k)_{k \in (\mathcal{S}_m\cup\mathcal{T}_n)\cap(\cup_{i\in [M]} \mathcal{S}_i)}\right) \\
&\overset{(\ref{message})}{=}&H\left((X_k)_{k \in (\mathcal{S}_m\cup\mathcal{T}_n)\cap(\cup_{i\in [M]} \mathcal{S}_i)}\big|(Z_k)_{k\in\mathcal{T}_n\setminus (\cup_{i\in [M]} \mathcal{S}_i)},(W_k)_{k \in (\mathcal{S}_m\cup\mathcal{T}_n)\cap(\cup_{i\in [M]} \mathcal{S}_i)}\right) \\
&=&H\left((X_k)_{k \in (\mathcal{S}_m\cup\mathcal{T}_n)\cap(\cup_{i\in [M]} \mathcal{S}_i)}\big|(Z_k)_{k\in\mathcal{T}_n\setminus (\cup_{i\in [M]} \mathcal{S}_i)}\right) \notag\\
&&-~I\left((X_k)_{k \in (\mathcal{S}_m\cup\mathcal{T}_n)\cap(\cup_{i\in [M]} \mathcal{S}_i)};(W_k)_{k \in (\mathcal{S}_m\cup\mathcal{T}_n)\cap(\cup_{i\in [M]} \mathcal{S}_i)}\big|(Z_k)_{k\in\mathcal{T}_n\setminus (\cup_{i\in [M]} \mathcal{S}_i)}\right) \\
&\geq&H\left((X_k)_{k \in (\mathcal{S}_m\cup\mathcal{T}_n)\cap(\cup_{i\in [M]} \mathcal{S}_i)}\big|(Z_k)_{k\in\mathcal{T}_n\setminus (\cup_{i\in [M]} \mathcal{S}_i)}\right) \notag\\
&&-~I\left((X_k)_{k \in [K]}; (W_{k})_{k\in\mathcal{T}_n \cap (\cup_{i\in [M]} \mathcal{S}_i)} \big|(Z_k)_{k\in\mathcal{T}_n\setminus (\cup_{i\in [M]} \mathcal{S}_i)}\right) \notag\\
&&-~ I\left((X_k)_{k \in [K]};(W_k)_{k\in\mathcal{S}_m} \big| (Z_k)_{k\in\mathcal{T}_n\setminus (\cup_{i\in [M]} \mathcal{S}_i)},(W_k)_{k\in \mathcal{T}_n \cap(\cup_{i\in [M]} \mathcal{S}_i)}\right) \label{eq:tq}\\
&\overset{(\ref{lemma1_eq})(\ref{ind})}{\geq}& \big|(\mathcal{S}_m\cup\mathcal{T}_n)\cap\left(\cup_{i\in [M]}\mathcal{S}_i\right)\big| L \notag\\
&&-~I\left((X_k)_{k \in [K]}; (W_{k})_{k\in\mathcal{T}_n \cap (\cup_{i\in [M]} \mathcal{S}_i)} \Bigg| \sum_{k\in[K]} W_k, (W_k, Z_k)_{k\in\mathcal{T}_n\setminus (\cup_{i\in [M]} \mathcal{S}_i)}\right) \notag\\
&&-~ I\left((X_k)_{k \in [K]};(W_k)_{k\in\mathcal{S}_m} \Bigg| \sum_{k\in[K]} W_k, (W_k, Z_k)_{k\in\mathcal{T}_n} \right) 
\label{eq:tq1}\\
&\overset{(\ref{security})}{=}& \big|(\mathcal{S}_m\cup\mathcal{T}_n)\cap\left(\cup_{i\in [M]}\mathcal{S}_i\right)\big| L
\end{eqnarray}
where (\ref{eq:tq}) follows from expanding the mutual information term with chain-rule and the property that $(\mathcal{S}_m\cup\mathcal{T}_n)\cap(\cup_{i\in [M]}\mathcal{S}_i) = \left(\mathcal{T}_n \cap (\cup_{i\in [M]} \mathcal{S}_i)\right) \cup \mathcal{S}_m$ (remember that $\mathcal{S}_m \cap \mathcal{T}_n = \emptyset$). In (\ref{eq:tq1}), the first term follows from chain-rule and applying Lemma \ref{lemma1} to each expanded term; the second term follows from adding corresponding conditional terms, the independence of $X_k$ and $Z_k$, and the property that $|\mathcal{S}_m \cup \mathcal{T}_n| \leq K-1$, then we can apply chain-rule to expand the mumtual information term to that of each term $W_k$ in the set $\mathcal{T}_n \cap \left(\cup_{i\in [M]} \mathcal{S}_i\right)$, where $k$ belongs to some explicit security input set and the conditioning term belongs to the colluding user set $\mathcal{T}_n$ so that the term is zero due to (\ref{security}); the third term is also zero, by applying the security constraint (\ref{security}) with $\mathcal{S}_m$ and $\mathcal{T}_n$.

\hfill\QED

Generalizing the above lemma to also include the implicit security input set $\mathcal{S}_I$, we have the following lemma.

\begin{lemma} \label{lemma4} 
For any $\mathcal{S}_m, \mathcal{T}_n, m\in [M],n\in [N]$ such that $\mathcal{S}_m \cap \mathcal{T}_n = \emptyset$ and $|\mathcal{S}_m\cup\mathcal{T}_n| \leq K-1$, we have 
\begin{eqnarray}
    H\left((Z_k)_{k \in (\mathcal{S}_m\cup\mathcal{T}_n)\cap\overline{\mathcal{S}}}\big|(Z_k)_{k\in\mathcal{T}_n\setminus \overline{\mathcal{S}}}\right) \geq \big|(\mathcal{S}_m\cup\mathcal{T}_n)\cap\overline{\mathcal{S}}\big| L. \label{lemma4_eq}
\end{eqnarray}
\end{lemma}

\proof Recall that $\overline{\mathcal{S}} = \cup_{i\in [M]} \mathcal{S}_i \cup \mathcal{S}_I$.
\begin{eqnarray}
&&H\left((Z_k)_{k \in (\mathcal{S}_m\cup\mathcal{T}_n)\cap(\cup_{i\in [M]} \mathcal{S}_i\cup\mathcal{S}_I)}\big|(Z_k)_{k\in\mathcal{T}_n\setminus (\cup_{i\in [M]} \mathcal{S}_i\cup\mathcal{S}_I)}\right)\notag\\
&=&H\left((Z_k)_{k \in (\mathcal{T}_n\cap\mathcal{S}_I)}\big|(Z_k)_{k\in\mathcal{T}_n\setminus (\cup_{i\in [M]} \mathcal{S}_i\cup\mathcal{S}_I)}\right)\notag\\
&&+~H\left((Z_k)_{k \in \left((\mathcal{S}_m\cup\mathcal{T}_n)\cap(\cup_{i\in [M]} \mathcal{S}_i\cup\mathcal{S}_I)\right)\setminus(\mathcal{T}_n\cap\mathcal{S}_I)}\big|(Z_k)_{k\in\mathcal{T}_n\setminus \overline{\mathcal{S}}},(Z_k)_{k \in (\mathcal{T}_n\cap\mathcal{S}_I)}\right) \\
&\geq&|\mathcal{T}_n\cap\mathcal{S}_I |L+\big|\left((\mathcal{S}_m\cup\mathcal{T}_n)\cap(\cup_{i\in [M]} \mathcal{S}_i\cup\mathcal{S}_I)\right)\setminus(\mathcal{T}_n\cap\mathcal{S}_I)\big|L \label{eq:tx} \\
&=&|\mathcal{T}_n\cap\mathcal{S}_I|L+|(\mathcal{S}_m\cup\mathcal{T}_n)\cap(\cup_{i\in [M]} \mathcal{S}_i\cup\mathcal{S}_I)|L\\
&&-~|(\mathcal{S}_m\cup\mathcal{T}_n)\cap(\cup_{i\in [M]} \mathcal{S}_i\cup\mathcal{S}_I)\cap(\mathcal{T}_n\cap\mathcal{S}_I)|L\\
&=&|\mathcal{T}_n\cap\mathcal{S}_I|L+|(\mathcal{S}_m\cup\mathcal{T}_n)\cap(\cup_{i\in [M]} \mathcal{S}_i\cup\mathcal{S}_I)|L-|\mathcal{T}_n\cap\mathcal{S}_I|L\\
&=&\big|(\mathcal{S}_m\cup\mathcal{T}_n)\cap\overline{\mathcal{S}}\big|L
\end{eqnarray}
where (\ref{eq:tx}) is derived as follows. First, consider the first term. For each element $u \in \mathcal{T}_n \cap \mathcal{S}_I$, as $u \in \mathcal{S}_I$, we have $\mathcal{S}_{m'}$ and $\mathcal{T}_{n'}$ such that $|\mathcal{S}_{m'} \cap \mathcal{T}_{n'}| = K-1$ and $u \cup \mathcal{S}_{m'} \cup \mathcal{T}_{n'} = [K]$.
\begin{eqnarray}
    &&H\left((Z_k)_{k \in \mathcal{T}_n\cap\mathcal{S}_I}\big|(Z_k)_{k\in\mathcal{T}_n\setminus (\cup_{i\in [M]} \mathcal{S}_i\cup\mathcal{S}_I)}\right) \notag\\
    &\geq&\sum_{u: u\in \mathcal{T}_n\cap\mathcal{S}_I}H\left(Z_{u}\big|(Z_k)_{k\in\mathcal{T}_n\setminus (\cup_{i\in [M]} \mathcal{S}_i\cup\mathcal{S}_I)},(Z_k)_{k\in (\mathcal{T}_n\cap\mathcal{S}_I)\setminus\{u\}}\right) \\
    &\geq&\sum_{u: u \in \mathcal{T}_n\cap\mathcal{S}_I}H\left(Z_{u}\big|(Z_k)_{k\in[K]\setminus (\cup_{i\in [M]} \mathcal{S}_i\cup\{u\})}\right)\\
    &\geq&\sum_{u: u \in \mathcal{T}_n\cap\mathcal{S}_I}H\left(Z_{u}\big|(Z_k)_{k\in[K]\setminus (\mathcal{S}_{m'}\cup\{u\})}\right)\\
    &=&\sum_{u: u \in \mathcal{T}_n\cap\mathcal{S}_I}H\left(Z_{u} \big|(Z_k)_{k\in\mathcal{T}_{n'}}\right)\\
    &\overset{(\ref{corollary1_eq})}{\geq}&|\mathcal{T}_n\cap\mathcal{S}_I|L.
\end{eqnarray}
Second, consider the second term of (\ref{eq:tx}).
\begin{eqnarray}
    &&H\left((Z_k)_{k \in \left((\mathcal{S}_m\cup\mathcal{T}_n)\cap(\cup_{i\in [M]} \mathcal{S}_i\cup\mathcal{S}_I)\right)\setminus(\mathcal{T}_n\cap\mathcal{S}_I)}\big|(Z_k)_{k\in\mathcal{T}_n\setminus \overline{\mathcal{S}}},(Z_k)_{k \in (\mathcal{T}_n\cap\mathcal{S}_I)}\right) \notag\\
    &=&H\left((Z_k)_{k \in \left((\mathcal{S}_m\cup\mathcal{T}_n)\cap(\cup_{i\in [M]} \mathcal{S}_i\cup\mathcal{S}_I)\right)\setminus(\mathcal{T}_n\cap\mathcal{S}_I)}\big|(Z_k)_{k\in\mathcal{T}_n\setminus (\cup_{i\in [M]} \mathcal{S}_i\cup\mathcal{S}_I)},(Z_k)_{k \in (\mathcal{T}_n\cap\mathcal{S}_I)}\right) \\
    &=&H\left((Z_k)_{k \in \left((\mathcal{S}_m\cup\mathcal{T}_n)\cap(\cup_{i\in [M]} \mathcal{S}_i\cup\mathcal{S}_I)\right)\setminus(\mathcal{T}_n\cap\mathcal{S}_I)}\big|(Z_k)_{k\in\mathcal{T}_n\setminus (\cup_{i\in [M]} \mathcal{S}_i)}\right) \\
    &\overset{(\ref{lemma3_eq})}{\geq}&\big|\left((\mathcal{S}_m\cup\mathcal{T}_n)\cap(\cup_{i\in [M]} \mathcal{S}_i\cup\mathcal{S}_I)\right)\setminus(\mathcal{T}_n\cap\mathcal{S}_I)\big|L
    \label{lemma4_eq3}
\end{eqnarray}
where (\ref{lemma4_eq3}) follows from the fact that $\left((\mathcal{S}_m\cup\mathcal{T}_n)\cap(\cup_{i\in [M]} \mathcal{S}_i\cup\mathcal{S}_I)\right)\setminus(\mathcal{T}_n\cap\mathcal{S}_I)= (\mathcal{S}_m\cup\mathcal{T}_n)\cap(\cup_{i\in [M]} \mathcal{S}_i)$.

\hfill\QED

We are now ready to show that $R_{Z_\Sigma} \geq \min(a^*, K-1)$. 
First, Lemma \ref{lemma4} directly gives us $R_{Z_\Sigma} \geq a^*$. Consider any $\mathcal{A}_{m,n}$ whose $\mathcal{S}_m, \mathcal{T}_n$ satisfy the conditions in Lemma \ref{lemma4} (note that the set systems are monotone so that we may assume without loss that $\mathcal{S}_m \cap \mathcal{T}_n = \emptyset$).
\begin{eqnarray}
    H(Z_{\Sigma})&\overset{(\ref{total rand})}{\geq}&H\left((Z_k)_{k\in [K]}\right)\\
    &\geq&H\left((Z_k)_{k \in (\mathcal{S}_{m}\cup\mathcal{T}_{n})\cap\overline{\mathcal{S}}}\big|(Z_k)_{k\in\mathcal{T}_{n}\setminus \overline{\mathcal{S}}}\right) \\
    &\overset{(\ref{lemma4_eq})}{\geq}& \big|(\mathcal{S}_{m}\cup\mathcal{T}_{n})\cap\overline{\mathcal{S}}\big| L = | \mathcal{A}_{m,n}| L\\
\Rightarrow ~~H(Z_{\Sigma}) &\geq& \mbox{max}_{m,n} |\mathcal{A}_{m,n}| L =a^*L\label{totalrand_eq1}\\
\Rightarrow ~~~~~~R_{Z_{\Sigma}} &=& \frac{L_{Z_{\Sigma}}}{L} \geq \frac{H(Z_{\Sigma})}{L} \geq a^*.
\end{eqnarray}

Second, a slight twists shows that $R_{Z_\Sigma} \geq K-1$ when $a^* = K$, i.e., there exist $\mathcal{S}_m, \mathcal{T}_n$ so that $|\mathcal{S}_m\cup\mathcal{T}_n|=K$. As $(\mathcal{S}_m)_m, (\mathcal{T}_n)_n$ are monotone, there exists $\mathcal{S}_{m'}, \mathcal{T}_n$ so that $|\mathcal{S}_{m'}\cup\mathcal{T}_{n}|=K-1$ and $\big|(\mathcal{S}_{m'}\cup\mathcal{T}_{n})\cap\overline{\mathcal{S}}\big|=K-1$. Applying Lemma \ref{lemma4}, we have 
\begin{eqnarray}
H(Z_{\Sigma})&\overset{(\ref{total rand})}{\geq}&H\left((Z_k)_{k\in [K]}\right)\geq H\left((Z_k)_{k \in (\mathcal{S}_{m'}\cup\mathcal{T}_{n})\cap\overline{\mathcal{S}}}\big|(Z_k)_{k\in\mathcal{T}_{n}\setminus \overline{\mathcal{S}}}\right) \\
&\overset{(\ref{lemma4_eq})}{\geq}& \big|(\mathcal{S}_{m'}\cup\mathcal{T}_{n})\cap\overline{\mathcal{S}}\big| L = (K-1)L\\
\Rightarrow ~~~ R_{Z_{\Sigma}} &=& \frac{L_{Z_{\Sigma}}}{L} \geq \frac{H(Z_{\Sigma})}{L} \geq K-1.
\end{eqnarray}

\subsection{Proof of Example \ref{ex2}}
Next we consider the `if' case of Theorem \ref{thm:d1}, where we need one more step to further tighten the bound $R_{Z_\Sigma} \geq a^*$ to include an additional term $b^*$. To appreciate the idea in a simpler setting, let's again start with an example.

\begin{example}\label{ex2}
Consider $K=5$, the security input sets are $(\mathcal{S}_1,\mathcal{S}_2,\mathcal{S}_3) =(\emptyset, \{1\}, \{2\})$, and the colluding user sets are $(\mathcal{T}_1,\cdots,\mathcal{T}_{9})=(\emptyset, \{1\}, \{2\}, \{3\}, \{4\},$ $\{5\}, \{1,3\}, \{2,4\}, \{2,5\})$. 
\end{example}

For Example \ref{ex2}, the implicit security set $\mathcal{S}_I$ is empty, and $\overline{\mathcal{S}}= \cup_{m} \mathcal{S}_m =\{1,2\}$. Finding all $\mathcal{A}_{m,n}$ with maximum cardinality, we have $\mathcal{A}_{2,3} = \mathcal{A}_{3,2} = \mathcal{A}_{2,8}  = \mathcal{A}_{2,9} = \mathcal{A}_{3,7} = \overline{\mathcal{S}} =  \{1,2\}$, and $a^* = \big|\overline{\mathcal{S}}\big| = 2 \leq K-1 = 4$. Then $\mathcal{Q} = \cup_{m,n: |\mathcal{A}_{m,n}| = a^*} \mathcal{S}_m \cup \mathcal{T}_n = \{1,2,3,4,5\}$ and $|\mathcal{Q}| = 5 = K$. So we are in the `if' case.

Consider all $\mathcal{A}_{m,n}$ sets so that $|\mathcal{A}_{m,n}| = a^*$. For Example \ref{ex2}, we have $5$ such sets $\mathcal{A}_{2,3}$, $\mathcal{A}_{3,2}$, $\mathcal{A}_{2,8}$, $\mathcal{A}_{2,9}$, $\mathcal{A}_{3,7}$. Note that $\mathcal{S}_m \cup \mathcal{T}_n = \mathcal{A}_{m,n} \cup (\mathcal{T}_n \setminus \overline{\mathcal{S}})$, and we consider the key variables $Z_k$ in the set $\mathcal{S}_m \cup \mathcal{T}_n$ and split them to $\mathcal{A}_{m,n}$ (treated by Lemma \ref{lemma4} and corresponds to $a^*$) and $\mathcal{T}_n \setminus \overline{\mathcal{S}}$ (the new part treated by a linear program and corresponds to $b^*$).
\begin{eqnarray}
H(Z_\Sigma) &\geq& \max\Big( H\left( \left(Z_k\right)_{k \in \mathcal{S}_2 \cup \mathcal{T}_3} \right), H\left( \left(Z_k\right)_{k \in \mathcal{S}_3 \cup \mathcal{T}_2} \right), H\left( \left(Z_k\right)_{k \in \mathcal{S}_2 \cup \mathcal{T}_8} \right), \notag\\
&&~~~~~~  H\left( \left(Z_k\right)_{k \in \mathcal{S}_2 \cup \mathcal{T}_9} \right), H\left( \left(Z_k\right)_{k \in \mathcal{S}_3 \cup \mathcal{T}_7} \right) \Big) \\
&=& \max \big( H(Z_1, Z_2), H(Z_1, Z_2, Z_4), H(Z_1, Z_2, Z_5), H(Z_1, Z_2, Z_3) \big) \\
&=& \max \big( H(Z_4) + H(Z_1, Z_2 | Z_4), H(Z_5) + H(Z_1, Z_2 | Z_5), H(Z_3) + H(Z_1, Z_2 | Z_3) \big) \label{eq:tt} \\
&\overset{(\ref{lemma4_eq})}{\geq}& \max \big( H(Z_4), H(Z_5), H(Z_3) \big) + a^*L \label{eq:tt1}
\end{eqnarray}
where in (\ref{eq:tt}), we split the non-redundant $Z_k$ term in set $\mathcal{S}_m \cup \mathcal{T}_n$ to that in $\mathcal{T}_n \setminus \overline{\mathcal{S}}$ and $\mathcal{A}_{m,n}$; in (\ref{eq:tt1}), we use Lemma \ref{lemma4} to bound the $\mathcal{A}_{m,n}$ term conditioned on $\mathcal{T}_n \setminus \overline{\mathcal{S}}$ (note that we consider only $\mathcal{A}_{m,n}$ where $|\mathcal{A}_{m,n}| = a^*$, i.e., $\mathcal{A}_{m,n} = \overline{\mathcal{S}}$). 
Next, we proceed to bound the term $\max \left( H(Z_4), H(Z_5), H(Z_3) \right)$, where it turns out that the only constraints required are from Lemma \ref{lemma2}. From (\ref{lemma2_eq}), we have
\begin{eqnarray}
    \mathcal{S}_2, \mathcal{T}_8: && H(Z_3,Z_5|Z_2,Z_4)\geq L, \notag \\
    \mathcal{S}_2, \mathcal{T}_9: && H(Z_3,Z_4|Z_2,Z_5)\geq L, \notag \\
    \mathcal{S}_3, \mathcal{T}_7: && H(Z_4,Z_5|Z_1,Z_3)\geq L. \label{eq:tt3}
\end{eqnarray}
To bound (\ref{eq:tt1}) with constraints in (\ref{eq:tt3}), we resort to a linear program, with conditional entropy terms as the variables that are consistent with chain-rule expansion. In particular, set $H(Z_3) = b_3L, H(Z_4|Z_3) = b_4L, H(Z_5|Z_3,Z_4) = b_5L$, then
\begin{eqnarray}
R_{Z_\Sigma} &\geq& a^* + \min \max(b_3, b_4, b_5)
\end{eqnarray}
where $\min$ is over the following linear constraints,
\begin{eqnarray}
    \mathcal{S}_2, \mathcal{T}_8: && b_3 +b_5 \geq \big(H(Z_3|Z_2,Z_4)+H(Z_5|Z_2,Z_3,Z_4)\big)/L \geq 1, \notag\\
    \mathcal{S}_2, \mathcal{T}_9: && b_3 +b_4 \geq \big(H(Z_3|Z_2,Z_5)+H(Z_4|Z_2,Z_3,Z_5)\big)/L \geq 1, \notag\\
    \mathcal{S}_3, \mathcal{T}_7: && b_4 +b_5 \geq \big(H(Z_4|Z_1,Z_3)+H(Z_5|Z_1,Z_3,Z_4)\big)/L \geq 1, \notag\\
    && b_3, b_4, b_5 \geq 0. \label{eq:tt4}
\end{eqnarray}
Intuitively, the correlation/conflict between $H(Z_3), H(Z_4), H(Z_5)$ is captured through (\ref{eq:tt3}) and we wish to find the tightest bound subject to (\ref{eq:tt3}) (interestingly and somewhat surprisingly, this bound turns out to be tight as we show in Section \ref{sec:ach}).
Therefore we have transformed the $R_{Z_\Sigma}$ converse to a linear program on non-negative variables $b_3,b_4, b_5$, where each $\mathcal{A}_{m,n}$ set with maximum cardinality contributes one linear constraint (some redundant ones are removed in (\ref{eq:tt4})).
Now defining $b^* = \min\max(b_3, b_4, b_5)$ as the optimal value of the linear program subject to constraints (\ref{eq:tt4}), we have obtained the desired converse $R_{Z_\Sigma} \geq a^* + b^*$. For Example \ref{ex2}, the optimal value is $b^* = 1/2$, attained when $b_3=b_4=b_5=1/2$ and this will be useful for the achievability proof, presented in Section \ref{sec:ach}.

\subsection{Proof of $R_{Z_\Sigma} \geq a^* + b^*$}
Building upon the insights from Example \ref{ex2}, we present the general proof of $R_{Z_\Sigma} \geq a^* + b^*$ when the `if' condition holds, i.e., $a^*\leq K-1$, $a^*=\big|\overline{\mathcal{S}}\big|$, and $|\mathcal{Q}|= K$. 

Note that $a^* = \big|\overline{\mathcal{S}}\big|$ so that each $\mathcal{A}_{m,n}$ such that $|\mathcal{A}_{m,n}| = a^*$ must satisfy $\mathcal{A}_{m,n} = \overline{\mathcal{S}}$. Consider all such $\mathcal{A}_{m,n}$ sets (without loss of generality, we assume for each such $\mathcal{A}_{m,n}$, $\mathcal{S}_m \cap \mathcal{T}_n = \emptyset$ and $|\mathcal{S}_m \cup \mathcal{T}_n| \leq K-1$ as the set systems are monotone and this will not change the linear program (\ref{lp}), i.e., the dropped ones are redundant). Following a similar decomposition as that of Example \ref{ex2}, we have
\begin{eqnarray}
  H(Z_{\Sigma})&\overset{(\ref{total rand})}{\geq}& \max_{m,n: |\mathcal{A}_{m,n}| = a^*}  H\big((Z_k)_{k\in \mathcal{S}_m \cap \mathcal{T}_n}\big) \\
    &\geq&\max_{m,n: |\mathcal{A}_{m,n}| = a^*}  \left( H\big((Z_k)_{k\in \mathcal{T}_{n}\setminus\overline{\mathcal{S}}}\big) +H\big((Z_k)_{k \in (\mathcal{S}_{m}\cup\mathcal{T}_{n})\cap\overline{\mathcal{S}}}\big|(Z_k)_{k\in\mathcal{T}_{n}\setminus \overline{\mathcal{S}}}\big)  \right) \\
    &\overset{(\ref{lemma4_eq})}{\geq}& \max_{m,n: |\mathcal{A}_{m,n}| = a^*}  H\big((Z_k)_{k\in \mathcal{T}_{n}\setminus\overline{\mathcal{S}}}\big)  + a^*L \label{eq:con1} 
\end{eqnarray}
subject to the following constraints by Lemma \ref{lemma2},
\begin{eqnarray}
\forall m,n ~\mbox{where $|\mathcal{A}_{m,n}| = a^*$}:~H\left((Z_k)_{k \in [K] \setminus(\mathcal{S}_{m}\cup\mathcal{T}_{n})}\big|(Z_k)_{k\in\mathcal{T}_{n}}\right) &\overset{(\ref{lemma2_eq})}{\geq}& L. \label{eq:con2}
\end{eqnarray}
Next, following the steps of the proof of Example \ref{ex2}, we translate the inequality (\ref{eq:con1}) and the constraints (\ref{eq:con2}) to a linear program in $b_k$ variables, defined as follows.
\begin{eqnarray}
\forall k\in [K]\setminus\overline{\mathcal{S}}, ~~b_k \triangleq H\left(Z_{k}\big|(Z_{l})_{l\in[K]\setminus\overline{\mathcal{S}}, l < k}\right)/L. \label{assume_eq1}
\end{eqnarray}
Then normalizing (\ref{eq:con1}) by $L$ on both hand sides and expanding the entropy term in (\ref{eq:con1}) and (\ref{eq:con2}) by chain-rule with lexicographic order, we have
\begin{eqnarray}
R_{Z_\Sigma} &\geq& a^* + \min \max_{m,n: |\mathcal{A}_{m,n}| = a^*} \left( \sum_{k\in \mathcal{T}_n \setminus \overline{\mathcal{S}} } b_k\right) \\
\mbox{subject to}&& \sum_{k \in [K] \setminus (\mathcal{S}_m \cup \mathcal{T}_n)} b_k \geq 1, \forall m,n ~\mbox{such that}~|\mathcal{A}_{m,n}| = a^*, \\
&& b_k \geq 0, k \in [K]\setminus \overline{\mathcal{S}}.
\end{eqnarray}
Note that $\mathcal{A}_{m,n} = \overline{\mathcal{S}} \subset (\mathcal{S}_m \cup \mathcal{T}_n)$, so the chain-rule expansion above can be bounded by $b_k$ terms.
According to the linear program (\ref{lp}) whose optimal value is defined as $b^*$, we have obtained the desired converse bound $R_{Z_\Sigma} \geq a^* + b^*$.

\section{Achievability Proof of Theorem \ref{thm:d1}} \label{sec:ach}
We similarly start from the simpler `otherwise' case. Henceforth, we assume $a^* \leq K-1$ because otherwise $a^* = K$ we may apply the scheme in Fig.~\ref{fig:model} which achieves $R_{Z_\Sigma} = K-1$ and has been proved to be correct and secure for $\mathcal{S}_m = [K]$ (so for any other $\mathcal{S}_m$) and any $\mathcal{T}_n$ in Theorem 1 of \cite{Zhao_Sun_Summation}. We have the following two cases.


\subsection{Achievable Scheme of $R_{Z_\Sigma} = a^*$ for `Otherwise' Case where $a^* < \big|\overline{\mathcal{S}}\big|$}\label{sec:ach1}

In this case, every user in the (implicit and explicit) security input set $\overline{\mathcal{S}}$ will be assigned a key variable. 
We will operate over the field $\mathbb{F}_{q^B}$, i.e., we group $B$ symbols from $\mathbb{F}_q$ together and view them as an element from $\mathbb{F}_{q^B}$. For field size requirements that are necessary for security, we set $q^B > a^* \binom{|\overline{\mathcal{S}}|}{a^*}$, i.e., the integer $B$ is chosen to be no smaller than $\log_q\left(a^* \binom{|\overline{\mathcal{S}}|}{a^*}\right)$.
Consider $a^*$ i.i.d. uniform variables put in a column vector, ${\bf s} = (S_1; \cdots ;S_{a^*}) \in \mathbb{F}_{q^B}^{a^* \times 1}$ and set the key variables as 
 \begin{eqnarray}
 Z_\Sigma &=& {\bf s} \\
 Z_k &=& {\bf h}_k \times {\bf s}, \forall k \in \overline{\mathcal{S}}\notag\\
 Z_k &=& 0, \forall k \in [K]\setminus\overline{\mathcal{S}}\label{eq:c11}
 \end{eqnarray}
 where ${\bf h}_k \in \mathbb{F}_{q^B}^{1\times a^*}$ are chosen as follows (suppose $\overline{\mathcal{S}} = \{k_1, \cdots, k_{|\overline{\mathcal{S}}|}\}$)
 \begin{eqnarray}
&& \mbox{each element of}~{\bf h}_{k_1}, \cdots, {\bf h}_{k_{|\overline{\mathcal{S}}|-1}}~\mbox{is chosen uniformly and i.i.d. from}~\mathbb{F}_{q^B}, \notag\\
&& {\bf h}_{k_{|\overline{\mathcal{S}}|}} \triangleq -\left( {\bf h}_{k_1} + \cdots + {\bf h}_{k_{|\overline{\mathcal{S}}|-1}}\right) \label{eq:c12}
 \end{eqnarray}
 so that
 \begin{eqnarray}
 \sum_{k \in [K]} Z_k \overset{(\ref{eq:c11})}{=} \sum_{k \in \overline{\mathcal{S}}} Z_k \overset{(\ref{eq:c12})}{=} 0. \label{eq:c13}
 \end{eqnarray}
 Finally, set $L=B$ and the sent messages as
 \begin{eqnarray}
     X_k = W_k + Z_k, \forall k \in [K].\label{eq:m1}
 \end{eqnarray}
 Note that each symbol above is from $\mathbb{F}_{q^B}$ so that $L = B, L_{Z_\Sigma} = B a^*$ and the key rate achieved is $R_{Z_\Sigma} = L_{Z_\Sigma}/L = a^*$, as desired. Correctness is guaranteed by noting that $\sum_{k\in[K]} {X_k} \overset{(\ref{eq:c13})}{=} \sum_{k\in [K]} W_k$. The security proof is deferred to Section \ref{sec:security}, which relies on proving the existence of a realization of the randomly generated ${\bf h}_k$ vectors that have certain full rank property.

\begin{remark}
For this case and all cases presented in the following, the communication rate $L_X/L$ is 1 and is optimal (minimum, i.e., weak security does not incur additional communication cost).
\end{remark}

\subsection{Achievable Scheme of $R_{Z_\Sigma} = a^*$ for `Otherwise' Case where $a^* = \big|\overline{\mathcal{S}}\big|$, $|\mathcal{Q}| < K$}\label{sec:ach2}

In this case, every user in the total security input set $\overline{\mathcal{S}}$ and one user outside $\mathcal{Q}$ will be assigned a key variable.
 Pick $B$ so that $q^B > a^* \binom{|\overline{\mathcal{S}}|+1}{a^*} = a^*(a^*+1)$ and operate over $\mathbb{F}_{q^B}$.
 Consider $a^*$ i.i.d. uniform variables, ${\bf s} = (S_1;\cdots;S_{a^*}) \in \mathbb{F}_{q^B}^{a^* \times 1}$. Find any $u \in [K]\setminus\mathcal{Q}$ (which must exist as $|\mathcal{Q}| < K$ and $u \notin \overline{\mathcal{S}}$ as $a^* = \big|\overline{\mathcal{S}}\big|$ and $\overline{\mathcal{S}} \subset \mathcal{Q}$)
 and set the key variables as 
 \begin{eqnarray}
  Z_\Sigma &=& {\bf s} \\
 Z_k &=& {\bf h}_k \times {\bf s}, \forall k \in (\overline{\mathcal{S}} \cup \{u\}) \notag\\
 Z_k &=& 0, \forall k \in [K]\setminus(\overline{\mathcal{S}}\cup\{u\})\label{eq:c21}
 \end{eqnarray}
   where ${\bf h}_k \in \mathbb{F}_{q^B}^{1\times a^*}$ are chosen as follows (suppose $\overline{\mathcal{S}} = \{k_1, \cdots, k_{|\overline{\mathcal{S}}|}\}$)
 \begin{eqnarray}
&& \mbox{each element of}~{\bf h}_{k_1}, \cdots, {\bf h}_{k_{|\overline{\mathcal{S}}|}}~\mbox{is chosen uniformly and i.i.d. from}~\mathbb{F}_{q^B}, \notag\\
&& {\bf h}_{u} \triangleq -\left( {\bf h}_{k_1} + \cdots + {\bf h}_{k_{|\overline{\mathcal{S}}|}}\right) \label{eq:c22}
 \end{eqnarray}
 so that
 \begin{eqnarray}
 \sum_{k \in [K]} Z_k \overset{(\ref{eq:c21})}{=} \sum_{k \in (\overline{\mathcal{S}}\cup\{u\}) } Z_k \overset{(\ref{eq:c22})}{=} 0. \label{eq:c23}
 \end{eqnarray}
  Finally, set $L=B$ and the sent messages as
 \begin{eqnarray}
     X_k = W_k + Z_k, \forall k \in [K].\label{eq:m2}
 \end{eqnarray}
 The key rate achieved is $R_{Z_\Sigma} = a^*$. Correctness holds because $\sum_{k\in[K]} {X_k} \overset{(\ref{eq:c23})}{=} \sum_{k\in [K]} W_k$ and security proof is presented in a unified manner in Section \ref{sec:security}. 
 
\subsection{Achievable Scheme of $R_{Z_\Sigma} = a^* + b^*$ for `If' Case}\label{sec:ach3}

Consider the `if' case, where every user will be assigned some key variables, the amount of which is according to the optimal solution of the linear program (\ref{lp}). 
Denote the $b_k, k\in [K]\setminus\overline{\mathcal{S}}$ values that attain the optimal value $b^*$ of (\ref{lp}) are 
\begin{eqnarray}
    b_k=b_k^*=\frac{p_k}{\overline{q}}, \forall k\in [K]\setminus\overline{\mathcal{S}} \label{eq:bk}
\end{eqnarray}
where the linear program has rational coefficients so that the optimal solution is also rational, i.e., $p_k, \overline{q}$ are integers (and non-negative).
As a result,
\begin{eqnarray}
    \sum_{k\in [K]\setminus\overline{\mathcal{S}}}b_k^*= \frac{\sum_{k \in [K]\setminus\overline{\mathcal{S}}} p_k}{\overline{q}} \triangleq \frac{\overline{p}}{\overline{q}}. \label{eq:bk}
\end{eqnarray}
Pick $B$ so that $q^B > (a^* +b^*)\overline{q} \binom{K \overline{q}}{(a^* + b^*)\overline{q}}$ and operate over $\mathbb{F}_{q^B}$.
Consider $\overline{p}+(a^*-1)\overline{q}$ i.i.d. uniform variables, ${\bf s} = (S_1;\cdots;S_{\overline{p}+(a^*-1)\overline{q}}) \in \mathbb{F}_{q^B}^{(\overline{p}+(a^*-1)\overline{q}) \times 1}$ and set the key variables as (suppose $\overline{\mathcal{S}} = \{k_1, \cdots, k_{|\overline{\mathcal{S}}|}\}$)
 \begin{eqnarray}
  Z_\Sigma &=& {\bf s} \\
 {Z}_k &=& {\bf F}_k \times {\bf G}_k \times {\bf s}, \forall k \in [K]\setminus\overline{\mathcal{S}} \notag\\
 {Z}_k &=& {\bf H}_k \times {\bf s}, \forall k \in \overline{\mathcal{S}}
 \label{eq:c31} 
 \end{eqnarray}
 where each element of ${\bf F}_i \in \mathbb{F}_{q^B}^{\overline{q}\times p_k}$, ${\bf G}_i \in \mathbb{F}_{q^B}^{p_k\times 
 (\overline{p}+(a^*-1)\overline{q})}$, $i \in [K] \setminus \overline{\mathcal{S}}$, ${\bf H}_j \in \mathbb{F}_{q^B}^{\overline{q}\times 
 (\overline{p}+(a^*-1)\overline{q})}, j \in \{k_1, \cdots, k_{|\overline{\mathcal{S}}|-1}\}$ are drawn uniformly and i.i.d. from $\mathbb{F}_{q^B}$
 and
 \begin{eqnarray}
&&  {\bf H}_{k_{|\overline{\mathcal{S}}|}} = - \left( \sum_{i\in[K]\setminus \overline{\mathcal{S}}} {\bf F}_i \times {\bf G}_i + \sum_{j \in \overline{\mathcal{S}} \setminus \{k_{|\overline{\mathcal{S}}|}\} } {\bf H}_j \right) \label{eq:c32}\\
&\Rightarrow& \sum_{k\in[K]} Z_k \overset{(\ref{eq:c31})(\ref{eq:c32})}{=} 0. \label{eq:c33}
 \end{eqnarray}

Finally, set $L=B \overline{q}$, i.e., ${W_k}=(W_{k,1}; \cdots; W_{k,\overline{q}}) \in \mathbb{F}_{q^B}^{\overline{q} \times 1}$ and the sent messages as
 \begin{eqnarray}
     {X}_k = {W}_k + {Z}_k, \forall k \in [K].\label{eq:m3}
 \end{eqnarray}
Correctness is similarly guaranteed by taking $\sum_{k\in[K]} X_k$ and (\ref{eq:c33}). Before proceeding to the security proof, we note that 
the key rate achieved is $R_{Z_\Sigma} = L_{Z_\Sigma}/L = (\overline{p} +(a^*-1)\overline{q})/\overline{q} \overset{(\ref{eq:bk})}{=} a^* + \sum_{k\in[K] \setminus \overline{\mathcal{S}}} b_k^*-1 \overset{(\ref{eq:lp})}{=} a^* + b^*$, where the last step is based on a crucial property of the linear program (\ref{lp}), proved next. 
\begin{lemma}\label{lemma:lp}
For the linear program (\ref{lp}), its optimal value $b^*$ and optimal solution $b^*_k$ satisfy
\begin{eqnarray}
b^* = \sum_{k\in [K]\setminus\overline{\mathcal{S}}} b^*_k - 1. \label{eq:lp}
\end{eqnarray}
\end{lemma}

\proof
As we are in the `if' case, the sets we consider satisfy some properties that we now discuss and will be used.
\begin{eqnarray}
\forall m,n: |\mathcal{A}_{m,n}| = \big|(\mathcal{S}_m \cup \mathcal{T}_n) \cap \overline{\mathcal{S}}\big| = a^*, ~\mbox{we have}~ \overline{\mathcal{S}} \subset (\mathcal{S}_m \cup \mathcal{T}_n) ~\mbox{as}~ a^* = |\overline{\mathcal{S}}| ~\mbox{for the `if' case}. \label{eq:s1}
\end{eqnarray}
Then 
$\forall m,n: |\mathcal{A}_{m,n}| = a^*$, we can decompose $[K] \setminus \overline{\mathcal{S}}$ to the following two disjoint sets.
\begin{eqnarray}
[K] \setminus \overline{\mathcal{S}}  &=& \left( \left(\mathcal{S}_m \cup \mathcal{T}_n\right) \setminus \overline{\mathcal{S}} \right) \cup \Big( \big([K] \setminus (\mathcal{S}_m \cup \mathcal{T}_n)\big) \setminus \overline{\mathcal{S}} \Big) \\
&\overset{(\ref{eq:s1})}{=}& \left( \mathcal{T}_n \setminus \overline{\mathcal{S}} \right) \cup \big( [K] \setminus (\mathcal{S}_m \cup \mathcal{T}_n) \big) \label{eq:s2}
\end{eqnarray}
where the last step follows from the fact that $\mathcal{S}_m \subset \overline{\mathcal{S}}$ and $\overline{\mathcal{S}} \subset (\mathcal{S}_m \cup \mathcal{T}_n)$.

We are now ready to prove (\ref{eq:lp}). Rewrite the linear program (\ref{lp}) in an equivalent but more transparent form (through defining an auxliary variable, $b$ for the `max' objective function).
 \begin{eqnarray}
&& \min ~~~b \label{elp1} \\
 \text{subject to} 
    && \sum_{k\in \mathcal{T}_{n}\setminus\overline{\mathcal{S}}} b_{k} \leq b, \forall m,n ~\mbox{such that}~ |\mathcal{A}_{m,n}| = a^*, 
    \label{elp} \\
    && \sum_{k\in [K]\setminus (\mathcal{S}_m \cup \mathcal{T}_n)} b_{k} \geq 1,\forall m,n ~\mbox{such that}~ |\mathcal{A}_{m,n}| = a^*,
    \label{elp2}
    \\
    && b_{k}\geq 0, \forall k\in [K]\setminus\overline{\mathcal{S}}.    \label{elp3}
 \end{eqnarray}
First, we prove $b^* \leq \sum_{k\in[K]\setminus\overline{\mathcal{S}}} b^*_k - 1$. 
Suppose the optimal value $b^* = \min b$ is taken for $m_1, n_1$ for the constraint (\ref{elp}), i.e., $|\mathcal{A}_{m_1,n_1}| = a^*$ and without loss of generality, we assume $|\mathcal{S}_{m_1} \cup \mathcal{T}_{n_1}| \leq K-1$ (as the set systems are monotone and the objective function only depends on $\mathcal{T}_{n_1}$). Then
\begin{eqnarray}
b^* &=& \sum_{k\in \mathcal{T}_{n_1}\setminus\overline{\mathcal{S}}} b^*_{k} \label{eq:s3}\\
&\overset{(\ref{eq:s2})}{=}& \sum_{k\in [K]\setminus\overline{\mathcal{S}}} b^*_{k} - \sum_{k\in [K]\setminus (\mathcal{S}_{m_1} \cup \mathcal{T}_{n_1}) } b^*_{k} \\
&\overset{(\ref{elp2})}{\leq}& \sum_{k\in [K]\setminus\overline{\mathcal{S}}} b^*_{k} - 1.
\end{eqnarray}
Second, we prove $b^* \geq \sum_{k\in[K]\setminus\overline{\mathcal{S}}} b^*_k - 1$. For the above linear program, at least one of constraints (\ref{elp2}) must be tight (otherwise the binding constraints on $b_k^*$ are all from (\ref{elp3}), i.e., $b^*_k =0$, which violate (\ref{elp2})) and suppose it is for $m_2, n_2$, i.e., $|\mathcal{A}_{m_2,n_2}| = a^*$, $|\mathcal{S}_{m_2} \cup \mathcal{T}_{n_2}| \leq K-1$, and
\begin{eqnarray}
1 &=& \sum_{k\in [K]\setminus (\mathcal{S}_{m_2} \cup \mathcal{T}_{n_2})} b_{k}^* \\
&\overset{(\ref{eq:s2})}{=}& \sum_{k\in [K]\setminus\overline{\mathcal{S}}} b^*_{k} -  \sum_{k\in \mathcal{T}_{n_2} \setminus\overline{\mathcal{S}}} b^*_{k} \\
&\overset{(\ref{elp})}{\geq}& \sum_{k\in [K]\setminus\overline{\mathcal{S}}} b^*_{k} - b^*.
\end{eqnarray}
The proof of Lemma \ref{lemma:lp} is now complete.

 \hfill\QED

 \subsection{Proof of Security}\label{sec:security} 
Consider any set $\mathcal{S}_m, \mathcal{T}_n, m\in [M], n\in [N]$ so that\footnote{Recall that $a^* \leq K-1$, then $|\mathcal{S}_m\cup\mathcal{T}_n|$ cannot be $[K]$ because otherwise $\overline{\mathcal{S}} = [K]$ (as the set systems are monotone so that any element in $[K]\setminus \cup_m \mathcal{S}_m$ belongs to $\mathcal{S}_I$) and $a^* = K$.} $|\mathcal{S}_m\cup\mathcal{T}_n|\leq K-1$. We show that the security constraint (\ref{security}) is satisfied for all three cases above.
\begin{eqnarray}
&& I\left(\left(W_k\right)_{k \in \mathcal{S}_m}; \left(X_k\right)_{k \in [K]} \Bigg| \sum_{k \in [K]} W_k, \left(W_k, Z_k\right)_{k \in \mathcal{T}_n} \right)\notag\\
&\overset{(\ref{eq:m1})(\ref{eq:m2})(\ref{eq:m3})}{=}& H\left(\left(W_k+Z_k\right)_{k \in [K]} \Bigg| \sum_{k \in [K]} W_k, \left(W_k, Z_k\right)_{k \in \mathcal{T}_n} \right)\notag\\
&&-~H\left(\left(W_k+Z_k\right)_{k \in [K]} \Bigg| \sum_{k \in [K]} W_k, \left(W_k, Z_k\right)_{k \in \mathcal{T}_n},\left(W_k\right)_{k \in \mathcal{S}_m} \right)\label{eq_thm1_a4}\\
&=& H\left(\left(W_k+Z_k\right)_{k \in [K]} \Bigg| \sum_{k \in [K]} W_k, \left(W_k, Z_k\right)_{k \in \mathcal{T}_n} \right)\notag\\
&&~-H\left(\left(W_k+Z_k\right)_{k \in [K]\setminus\mathcal{T}_n} \Bigg| \sum_{k \in [K]} W_k, \left(W_k, Z_k\right)_{k \in \mathcal{T}_n},\left(W_k\right)_{k \in \mathcal{S}_m} \right)\label{eq_thm1_a5}\\
&=& H\left(\left(W_k+Z_k\right)_{k \in [K]} \Bigg| \sum_{k \in [K]} W_k, \left(W_k, Z_k\right)_{k \in \mathcal{T}_n} \right)\notag\\
&&-~H\left(\left(W_k+Z_k\right)_{k \in \mathcal{S}_m\setminus\mathcal{T}_n} \Bigg|  \sum_{k \in [K]} W_k, \left(W_k\right)_{k \in (\mathcal{S}_m\cup\mathcal{T}_n)},\left(Z_k\right)_{k \in \mathcal{T}_n}  \right)\notag\\
&&-~H\left(\left(W_k+Z_k\right)_{k \in [K]\setminus(\mathcal{S}_m\cup\mathcal{T}_n)} \Bigg| \sum_{k \in [K]} W_k, \left(W_k\right)_{k \in (\mathcal{S}_m\cup\mathcal{T}_n)},\left(Z_k\right)_{k \in (\mathcal{S}_m\cup\mathcal{T}_n)}\right)\\
&\overset{(\ref{ind})}{=}& H\left(\left(W_k+Z_k\right)_{k \in [K]} \Bigg| \sum_{k \in [K]} W_k, \left(W_k, Z_k\right)_{k \in \mathcal{T}_n} \right)-H\left(\left(Z_k\right)_{k \in \mathcal{S}_m\setminus\mathcal{T}_n} \Bigg|\left(Z_k\right)_{k \in \mathcal{T}_n}  \right) \notag\\
&&-~H\left(\left(W_k+Z_k\right)_{k \in [K]\setminus(\mathcal{S}_m\cup\mathcal{T}_n)} \Bigg| \sum_{k \in [K]} W_k, \left(W_k\right)_{k \in (\mathcal{S}_m\cup\mathcal{T}_n)},\left(Z_k\right)_{k \in (\mathcal{S}_m\cup\mathcal{T}_n)}\right)\label{security_eq1}\\
&\leq&|\mathcal{S}_m\setminus\mathcal{T}_n|L-|\mathcal{S}_m\setminus\mathcal{T}_n|L\\
&=&0
\end{eqnarray}
where in (\ref{security_eq1}), the difference of the first term and the third term is no greater than $|\mathcal{S}_m\setminus\mathcal{T}_n|L$, derived below and the second term is also $|\mathcal{S}_m\setminus\mathcal{T}_n|L$, proved in Lemma \ref{lemma:independent} below.
\begin{eqnarray}
    && H\left(\left(W_k+Z_k\right)_{k \in [K]} \Bigg| \sum_{k \in [K]} W_k, \left(W_k, Z_k\right)_{k \in \mathcal{T}_n} \right)\notag\\
    &&-~H\left(\left(W_k+Z_k\right)_{k \in [K]\setminus(\mathcal{S}_m\cup\mathcal{T}_n)} \Bigg| \sum_{k \in [K]} W_k, \left(W_k\right)_{k \in (\mathcal{S}_m\cup\mathcal{T}_n)},\left(Z_k\right)_{k \in (\mathcal{S}_m\cup\mathcal{T}_n)}\right)\\
    &=& H\left(\left(W_k+Z_k\right)_{k \in (\mathcal{S}_m\cup\mathcal{T}_n)} \Bigg| \sum_{k \in [K]} W_k, \left(W_k, Z_k\right)_{k \in \mathcal{T}_n} \right)\notag\\
    &&+~H\left(\left(W_k+Z_k\right)_{k \in [K]\setminus(\mathcal{S}_m\cup\mathcal{T}_n)} \Bigg| \sum_{k \in [K]} W_k, \left(W_k, Z_k\right)_{k \in \mathcal{T}_n},\left(W_k+Z_k\right)_{k \in (\mathcal{S}_m\cup\mathcal{T}_n)} \right)\notag\\
    &&-~H\left(\left(W_k+Z_k\right)_{k \in [K]\setminus(\mathcal{S}_m\cup\mathcal{T}_n)} \Bigg| \sum_{k \in [K]} W_k, \left(W_k\right)_{k \in (\mathcal{S}_m\cup\mathcal{T}_n)},\left(Z_k\right)_{k \in (\mathcal{S}_m\cup\mathcal{T}_n)}\right)\\
     &\leq& H\left(\left(W_k+Z_k\right)_{k \in (\mathcal{S}_m\setminus\mathcal{T}_n)} \right)\notag\\
     &&+~H\left(\left(W_k+Z_k\right)_{k \in [K]\setminus(\mathcal{S}_m\cup\mathcal{T}_n)} \Bigg| \sum_{k \in [K]} (W_k+Z_k), \left(W_k, Z_k\right)_{k \in \mathcal{T}_n},\left(W_k+Z_k\right)_{k \in (\mathcal{S}_m\cup\mathcal{T}_n)} \right)\notag\\
    &&-~H\left(\left(W_k+Z_k\right)_{k \in [K]\setminus(\mathcal{S}_m\cup\mathcal{T}_n)} \Bigg| \sum_{k \in [K]} (W_k+Z_k), \left(W_k\right)_{k \in (\mathcal{S}_m\cup\mathcal{T}_n)},\left(Z_k\right)_{k \in [K]}\right) \label{eq:ss1}\\
     &\overset{}{\leq}& H\left(\left(W_k+Z_k\right)_{k \in (\mathcal{S}_m\setminus\mathcal{T}_n)} \right)\notag\\
    &&+~H\left(\left(W_k+Z_k\right)_{k \in [K]\setminus(\mathcal{S}_m\cup\mathcal{T}_n)} \Bigg| \sum_{k \in [K]\setminus(\mathcal{S}_m\cup\mathcal{T}_n)} (W_k+Z_k)\right)\notag\\
    &&-~H\left(\left(W_k\right)_{k \in [K]\setminus(\mathcal{S}_m\cup\mathcal{T}_n)} \Bigg| \sum_{k \in [K]\setminus(\mathcal{S}_m\cup\mathcal{T}_n)} W_k, \left(W_k\right)_{k \in (\mathcal{S}_m\cup\mathcal{T}_n)},\left(Z_k\right)_{k \in [K]} \right) \label{eq:ss2}\\
    &\overset{(\ref{ind})(\ref{h2})}{\leq}& |\mathcal{S}_m\setminus\mathcal{T}_n|L + \left(\big|[K]\setminus(\mathcal{S}_m\cup\mathcal{T}_n)\big| - 1\right)L - \left(\big|[K]\setminus(\mathcal{S}_m\cup\mathcal{T}_n)\big| - 1\right)L \label{eq:ss3}\\
    &=& |\mathcal{S}_m\setminus\mathcal{T}_n|L
\end{eqnarray}
where to obtain the second and third term of (\ref{eq:ss1}), we use the property that the key variables are zero-sum for all cases, $\sum_{k \in [K]} Z_k = 0$ (refer to (\ref{eq:c13}), (\ref{eq:c23}), (\ref{eq:c33})) and adding conditioning cannot increase entropy. In (\ref{eq:ss3}), we bound the first two terms with the number of elements and the third term is due to the independence of $(W_k)_k$ (which is further uniform) and $(Z_k)_k$.

To complete the proof, we are left to prove the following lemma.
\begin{lemma}\label{lemma:independent}
For all cases of the achievable scheme presented in Section \ref{sec:ach1}, \ref{sec:ach2}, and \ref{sec:ach3}, we have
\begin{eqnarray}
H\left((Z_k)_{k \in \mathcal{S}_m\setminus\mathcal{T}_n} \big|(Z_k)_{k \in \mathcal{T}_n}  \right) = |\mathcal{S}_m\setminus\mathcal{T}_n|L, \forall m \in [M], n \in [N]. \label{eq:independent}
\label{eq:independent}
\end{eqnarray}
\end{lemma}

\interfootnotelinepenalty=10000
\proof We prove each case one by one. First, consider the scheme in Section \ref{sec:ach1}, where each user in $\overline{\mathcal{S}}$ is assigned a scalar key over $\mathbb{F}_{q^B}$ (refer to (\ref{eq:c11})). According to the design of the key coefficients ${\bf h}_k, k \in \overline{\mathcal{S}}$ (refer to (\ref{eq:c12})), we know that there exists a realization\footnote{The matrix formed by any $a^*$ distinct ${\bf h}_k$ vectors has a non-zero determinant polynomial (noting that $a^* < |\overline{\mathcal{S}}|$ for Section \ref{sec:ach1}). By Schwartz–Zippel lemma, the product of the determinant polynomials of the matrix formed by any $a^*$ ${\bf h}_k$ vectors has degree $a^* \binom{|\overline{\mathcal{S}}|}{a^*}$ and as the field size $q^B$ is chosen to be larger than that, the probability of the product polynomial to be non-zero is strictly positive, so we have the claim in (\ref{eq:sz1}).} of ${\bf h}_k, k \in \overline{\mathcal{S}}$ such that
\begin{eqnarray}
\mbox{any $a^*$ or fewer distinct ${\bf h}_k, k \in \overline{\mathcal{S}}$ vectors are linearly independent.}  \label{eq:sz1}
\end{eqnarray}
Equipped with (\ref{eq:sz1}), we are now ready to prove (\ref{eq:independent}). For any set $\mathcal{A} = \{a_1, \cdots, a_{|\mathcal{A}|}\}$, denote $[{\bf h}_k]_{k \in \mathcal{A}} \triangleq [{\bf h}_{a_1}; \cdots; {\bf h}_{a_{|\mathcal{A}|}}]$ as the row stack of the vectors.
\begin{eqnarray}
H\left((Z_k)_{k \in \mathcal{S}_m\setminus\mathcal{T}_n} |(Z_k)_{k \in \mathcal{T}_n}  \right) &=& H\big((Z_k)_{k \in \mathcal{S}_m \cup \mathcal{T}_n}\big) - H\left((Z_k)_{k \in \mathcal{T}_n}  \right) \\
&\overset{(\ref{eq:c11})}{=}& H\left((Z_k)_{k \in (\mathcal{S}_m \cup \mathcal{T}_n) \cap \overline{\mathcal{S}} } \right) - H\left((Z_k)_{k \in \mathcal{T}_n \cap \overline{\mathcal{S}} }  \right) \label{eq:st1}\\
&\overset{(\ref{eq:c11})}{=}& H\left([{\bf h}_k]_{k \in (\mathcal{S}_m \cup \mathcal{T}_n) \cap \overline{\mathcal{S}} } \times {\bf s}\right) - H\left([{\bf h}_k]_{k \in \mathcal{T}_n \cap \overline{\mathcal{S}} } \times {\bf s} \right) \\
&\overset{}{=}& \mbox{rank}\left([{\bf h}_k]_{k \in (\mathcal{S}_m \cup \mathcal{T}_n) \cap \overline{\mathcal{S}} } \right) L - \mbox{rank}\left([{\bf h}_k]_{k \in \mathcal{T}_n \cap \overline{\mathcal{S}} } \right) L \label{eq:st2} \\
&\overset{(\ref{eq:sz1})}{=}&  \big|(\mathcal{S}_m \cup \mathcal{T}_n) \cap \overline{\mathcal{S}}\big| L - \big| \mathcal{T}_n \cap \overline{\mathcal{S}} \big| L \label{eq:st3} \\
&=& \big|(\mathcal{S}_m\setminus\mathcal{T}_n) \cap \overline{\mathcal{S}} \big| L= |\mathcal{S}_m\setminus\mathcal{T}_n|L
\end{eqnarray}
where (\ref{eq:st1}) uses the fact that $Z_k = 0, k \in [K]\setminus\overline{\mathcal{S}}$ (see (\ref{eq:c11})). (\ref{eq:st2}) uses the fact that each element of ${\bf s}$ is uniform and independent, and we measure entropy in $q$-ary unit while the field operated is $\mathbb{F}_{q^B}$ and $B = L$. To obtain (\ref{eq:st3}), we use the property that $|\mathcal{A}_{m,n}| = \big|(\mathcal{S}_m \cup \mathcal{T}_n) \cap \overline{\mathcal{S}}\big| \leq a^*$ for Section \ref{sec:ach1} so that the ${\bf h}_k$ vectors are linearly independent (see (\ref{eq:sz1})). The last step follows from the fact that $\mathcal{S}_m \subset \overline{\mathcal{S}}$.

The proof for the other two cases is similar to that above so that we highlight the differences below. Second, consider the scheme in Section \ref{sec:ach2}, where each user in $\overline{\mathcal{S}} \cup \{u\}$ (and $u \notin \overline{\mathcal{S}}$) is assigned a scalar key over $\mathbb{F}_{q^B}$ (refer to (\ref{eq:c21})). According to the design of ${\bf h}_k, k \in \overline{\mathcal{S}} \cup \{u\}$ (refer to (\ref{eq:c22})) and the choice of the field size, by a similar reasoning with Schwartz–Zippel lemma, we know that there exists a realization 
of ${\bf h}_k, k \in \overline{\mathcal{S}} \cap \{u\}$ such that
\begin{eqnarray}
\mbox{any $a^*$ or fewer distinct ${\bf h}_k, k \in \overline{\mathcal{S}} \cup \{u\}$ vectors are linearly independent.}  \label{eq:sz2}
\end{eqnarray}
We now proceed to prove (\ref{eq:independent}). We have two sub-cases. 
For the first sub-case, $u \in \mathcal{T}_n$. Then $\big|(\mathcal{S}_m \cup \mathcal{T}_n) \cap \overline{\mathcal{S}}\big| < \big|\overline{\mathcal{S}}\big|= a^*$ because otherwise $\big|(\mathcal{S}_m \cup \mathcal{T}_n) \cap \overline{\mathcal{S}}\big| = \big|\overline{\mathcal{S}}\big| = a^*$ and according to Definition \ref{def:uni}, we have $u \in (\mathcal{S}_m \cup \mathcal{T}_n) \subset \mathcal{Q}$, which violates our choice of $u$ to be outside $\mathcal{Q}$ in Section \ref{sec:ach2}. 
\begin{eqnarray}
&& H\left((Z_k)_{k \in \mathcal{S}_m\setminus\mathcal{T}_n} |(Z_k)_{k \in \mathcal{T}_n}  \right) 
\notag\\
&\overset{(\ref{eq:c21})}{=}& H\left((Z_k)_{k \in (\mathcal{S}_m \cup \mathcal{T}_n) \cap (\overline{\mathcal{S}} \cup \{u\})} \right) - H\left((Z_k)_{k \in \mathcal{T}_n \cap (\overline{\mathcal{S}} \cup \{u\}) }  \right) \\
&\overset{}{=}& \mbox{rank}\left([{\bf h}_k]_{k \in (\mathcal{S}_m \cup \mathcal{T}_n) \cap (\overline{\mathcal{S}} \cup \{u\})} \right) L - \mbox{rank}\left([{\bf h}_k]_{k \in \mathcal{T}_n \cap (\overline{\mathcal{S}} \cup \{u\}) }  \right)L \\
&\overset{(\ref{eq:sz2})}{=}&  \big|(\mathcal{S}_m \cup \mathcal{T}_n) \cap (\overline{\mathcal{S}} \cup \{u\}) \big| L - \big| \mathcal{T}_n \cap ( \overline{\mathcal{S}} \cap \{u\}) \big| L \label{eq:st5} \\
&=& \big|(\mathcal{S}_m\setminus\mathcal{T}_n) \cap (\overline{\mathcal{S}} \cup \{u\})\big| L= |\mathcal{S}_m\setminus\mathcal{T}_n|L
\end{eqnarray}
where as $\big|(\mathcal{S}_m \cup \mathcal{T}_n) \cap \overline{\mathcal{S}}\big| < a^*$ , the first term of (\ref{eq:st5}) is no greater than $a^*$ enabling us to use (\ref{eq:sz2}) to reduce the rank to the set cardinality. The last step uses the fact that $u \in \mathcal{T}_n$.
For the second sub-case, $u \notin \mathcal{T}_n$ (recall that $u \notin \overline{\mathcal{S}}$ and $\mathcal{S}_m \subset \overline{\mathcal{S}}$).
\begin{eqnarray}
H\left((Z_k)_{k \in \mathcal{S}_m\setminus\mathcal{T}_n} |(Z_k)_{k \in \mathcal{T}_n}  \right)
&\overset{(\ref{eq:c21})}{=}& H\left((Z_k)_{k \in (\mathcal{S}_m \cup \mathcal{T}_n) \cap (\overline{\mathcal{S}} \cup \{u\})} \right) - H\left((Z_k)_{k \in \mathcal{T}_n \cap (\overline{\mathcal{S}} \cup \{u\}) }  \right) \\
&\overset{}{=}& H\left((Z_k)_{k \in (\mathcal{S}_m \cup \mathcal{T}_n) \cap \overline{\mathcal{S}} } \right) - H\left((Z_k)_{k \in \mathcal{T}_n \cap \overline{\mathcal{S}} }  \right) \label{eq:st4}\\
&\overset{(\ref{eq:sz2})}{=}&  \big|(\mathcal{S}_m \cup \mathcal{T}_n) \cap \overline{\mathcal{S}}\big| L - \big| \mathcal{T}_n \cap \overline{\mathcal{S}} \big| L 
= |\mathcal{S}_m\setminus\mathcal{T}_n|L
\end{eqnarray}
where (\ref{eq:st4}) uses the assumption that $u \notin \mathcal{T}_n$.

Third, consider the scheme in Section \ref{sec:ach3}, where each user in $\overline{\mathcal{S}}$ is assigned a key of entropy/rank $\overline{q}$ over $\mathbb{F}_{q^B}$ and each user in $[K]\setminus \overline{\mathcal{S}}$ is assigned a key of entropy $p_k$ over $\mathbb{F}_{q^B}$ (refer to (\ref{eq:c31})). According to the design of ${\bf H}_k, {\bf F}_k, {\bf G}_k$ (refer to (\ref{eq:c32}) and the paragraph above) and the choice of the field size, by a similar reasoning with Schwartz–Zippel lemma, we know that there exists a realization\footnote{Note that in total, for all ${\bf F}_k^1 \times {\bf G}_k, {\bf H}_k$, there are $\sum_{k \in [K]\setminus \overline{\mathcal{S}}} p_k + a^* \overline{q} \overset{(\ref{eq:bk})}{=} \sum_{k \in [K]\setminus \overline{\mathcal{S}}} b_k^* \overline{q} + a^* \overline{q} \overset{(\ref{eq:lp})}{=} (a^* + b^* + 1)\overline{q}$ row vectors. Then by the correlation of (\ref{eq:c32}) and the independent choices of ${\bf F}_k, {\bf G}_k, {\bf H}_k$, any $(a^* + b^*)\overline{q}$ rows 
can be freely set to a full rank matrix such that its determinant polynomial is not constantly zero.} 
of ${\bf H}_k, {\bf F}_k, {\bf G}_k$ such that (${\bf F}_k^1$ is defined as the first $p_k$ rows of ${\bf F}_k$)
\begin{eqnarray}
\mbox{the rows of ${\bf F}_k^1 \times {\bf G}_k, {\bf H}_k$ are linearly independent if they have at most $(a^*+b^*)\overline{q}$ rows.}  \label{eq:sz3}
\end{eqnarray}
We now proceed to prove (\ref{eq:independent}).
\begin{eqnarray}
&&H\left((Z_k)_{k \in \mathcal{S}_m\setminus\mathcal{T}_n} \big|(Z_k)_{k \in \mathcal{T}_n}  \right) \notag\\
&=& H\big((Z_k)_{k \in \mathcal{S}_m\cup\mathcal{T}_n} \big) - H\big( (Z_k)_{k \in \mathcal{T}_n}  \big)\\
&=& H\left((Z_k)_{k \in \left( (\mathcal{S}_m\cup\mathcal{T}_n) \cap \overline{\mathcal{S}}\right) \cup \left( (\mathcal{S}_m\cup\mathcal{T}_n) \setminus \overline{\mathcal{S}}\right) } \right) - H\left( (Z_k)_{k \in (\mathcal{T}_n \cap \overline{\mathcal{S}}) \cup (\mathcal{T}_n \setminus \overline{\mathcal{S}})}  \right) \\
&=& H\left((Z_k)_{k \in \left( (\mathcal{S}_m\cup\mathcal{T}_n) \cap \overline{\mathcal{S}}\right) \cup ( \mathcal{T}_n \setminus \overline{\mathcal{S}}) } \right) - H\left( (Z_k)_{k \in (\mathcal{T}_n \cap \overline{\mathcal{S}}) \cup (\mathcal{T}_n \setminus \overline{\mathcal{S}})}  \right) \\
&\overset{(\ref{eq:sz3})}{=}& \left(\big|(\mathcal{S}_m\cup\mathcal{T}_n) \cap \overline{\mathcal{S}}\big| \overline{q} + \sum_{k \in \mathcal{T}_n \setminus \overline{\mathcal{S}}} p_k \right) B - \left( \big|\mathcal{T}_n  \cap \overline{\mathcal{S}}\big| \overline{q} + \sum_{k \in \mathcal{T}_n \setminus \overline{\mathcal{S}}} p_k \right) B \label{eq:sf1} \\
&=& |\mathcal{S}_m\setminus\mathcal{T}_n| \overline{q}B = |\mathcal{S}_m\setminus\mathcal{T}_n| L
\end{eqnarray}
where $L = \overline{q}B$ for the scheme in Section \ref{sec:ach3} is used in the last step and in order to apply (\ref{eq:sz3}) to obtain (\ref{eq:sf1}), we note that $H\left((Z_k)_{k\in\mathcal{T}_n\setminus\overline{\mathcal{S}}}\right)$ is captured by the first $p_k$ rows (as ${\bf F}_k \times {\bf G}_k$ has rank $p_k$, refer to (\ref{eq:c31})) and need to show that the first term is no greater than $(a^* + b^*)\overline{q}$, the proof of which is provided for two sub-cases. For the first sub-case, $\big|(\mathcal{S}_{m}\cup\mathcal{T}_{n})\cap\overline{\mathcal{S}}\big|=a^*$, then $\big|(\mathcal{S}_m\cup\mathcal{T}_n) \cap \overline{\mathcal{S}}\big| \overline{q} + \sum_{k \in \mathcal{T}_n \setminus \overline{\mathcal{S}}} p_k = a^* \overline{q} + \overline{q} \sum_{k \in \mathcal{T}_n \setminus \overline{\mathcal{S}}} b_k^* \leq (a^* + b^*)\overline{q}$, where the last step follows from the `max' objective function of the linear program (\ref{lp}).
For the second sub-case, $\big|(\mathcal{S}_{m}\cup\mathcal{T}_{n})\cap\overline{\mathcal{S}}\big| < a^*$,
then $\big|(\mathcal{S}_m\cup\mathcal{T}_n) \cap \overline{\mathcal{S}}\big| \overline{q} + \sum_{k \in \mathcal{T}_n \setminus \overline{\mathcal{S}}} p_k \leq (a^*-1) \overline{q} + \overline{q} \sum_{k \in [K] \setminus \overline{\mathcal{S}}} b_k^* \overset{(\ref{eq:lp})}{=} (a^* -1 + b^* + 1)\overline{q}$ = $(a^*+b^*)q$.

\hfill \QED

\section{Discussion}
In this work, we have characterized the fundamental limits of weakly secure summation with arbitrary security constraints (where the weak security notion is similar to and a generalization of that considered in the network coding context \cite{Bhattad_Krishna, Silva_Kschischang, Yan_Sprintson_Zelenko}), and arbitrary colluding constraints (similar to those in private information retrieval \cite{Yao_Liu_Kang_Arbitrary, Cheng_Liu_Kang_Li}). As the security and colluding constraints can be arbitrarily heterogeneous, it turns out that interestingly, their interaction can be captured by a linear program with a number of linear constraints that on the one hand, impose the security constraint for each security input and colluding user set and on the other hand, attempt to minimize the key size (the max objective function in (\ref{lp}) can be transformed to constraints on the additional key consumption, refer to (\ref{elp2})). The resolving of such tension gives rise to matching converse claim and achievability argument (connected by the crucial algebraic property of the linear program in Lemma \ref{lemma:lp}) so that the exact information theoretic answer is obtained.

Going forward, we remind that our model requires the security constraint (\ref{security}) to be satisfied for {\em each} security input set and {\em each} colluding user set (i.e., a product model), which can be further relaxed to an individual pair model, i.e., when certain colluding user set is present, the security input set can only take one choice instead of all security input sets. In other words, the allowed pairs of colluding user set and security input set are specified and are not all products. We note that this model can be more general than the one we considered and our techniques do not appear to be sufficient to address this generalized model, which is left as an interesting future work.

More broadly, secure summation is an information theoretic primitive whose model can be further enriched to catch new requirements in federated learning, e.g., user dropout \cite{aggregation, aggregation_turbo, aggregation_fast, Zhao_Sun_Aggregate, aggregation_light, aggregation_swift, WSJC_Groupwise, Wang_Ulukus_FSL}, user selection \cite{Zhao_Sun_MDS, cho2020client, mohamed2021privacy, wang2022unified}, groupwise keys \cite{WSJC_Groupwise, Zhao_Sun_Summation} etc. Considerations of weak security constraints in these settings are promising directions for novel insights.

\let\url\nolinkurl
\bibliographystyle{IEEEtran}
\bibliography{Thesis}
\end{document}